\documentclass[letterpaper,preprint, amsmath,amssymb, aps, prd ]{revtex4-1}

\usepackage{graphicx}
\usepackage{dcolumn}
\usepackage{bm}
\usepackage{overpic}
\usepackage{hyperref}

\usepackage{atlasphysics}
\usepackage{wptdefs}

\begin{document}

\preprint{CERN-PH-EP-2011-134}
\preprint{Submitted to Physical Review D}

\vspace*{0.1cm}

\title{Measurement of the Transverse Momentum Distribution of $W$ Bosons in 
$pp$ Collisions at $\sqrt{s} =7$~TeV with the ATLAS Detector}

\author{The ATLAS Collaboration}

\date{November 3, 2011}

\begin{abstract}
This paper describes a measurement of the $W$ boson transverse momentum distribution using 
ATLAS $pp$ collision data from the 2010 run of the LHC at $\sqrt{s} = 7$~\TeV, 
corresponding to an integrated luminosity of about 31~\ipb. Events from 
both \Wen\ and \Wmn\ are used, and the transverse momentum of 
the $W$ candidates is measured through the energy deposition in the calorimeter 
from the recoil of the $W$.  The resulting distributions are unfolded to obtain 
the normalized differential cross sections as a function of the $W$ boson 
transverse momentum. We present results for $\ptw < 300$~GeV in
the electron and muon channels  as well as for their combination, and
compare the combined results  to the predictions of perturbative QCD
and a selection of event generators. 
\end{abstract}

\maketitle

\section{\label{sec:Introduction}Introduction}

At hadron colliders, $W$ and $Z$ bosons are produced with non-zero 
momentum transverse to the beam direction due to parton radiation from the initial
state.  Measuring the transverse
momentum ($p_T$) distributions of $W$ and $Z$ bosons at the LHC provides a useful test of QCD calculations,
because different types of calculations are expected to produce the most 
accurate predictions for the low-$p_T$ and high-$p_T$ parts of the spectrum.  
This measurement complements studies which constrain the proton parton distribution
functions (PDFs), such as the $W$ lepton charge 
asymmetry in $pp$ collisions~\cite{WChargeAsymPaper}, because the dynamics which generate
transverse momentum in the $W$ do not depend strongly on the distribution of the proton
momentum among the partons.
The $W$ $p_T$ is reconstructed in \Wln\ events (where $\ell = e$ or $\mu$ in this paper).
Because of the neutrino in the final state, the $W$ $p_T$ must be reconstructed 
through the hadronic recoil, which is the energy observed in the 
calorimeter excluding the lepton signature.  
This measurement is therefore also complementary to measurements of the $Z$ $p_T$, 
which is measured using $\Zll$ events in which the $Z$ $p_T$ is reconstructed 
via the momentum of the lepton pair~\cite{ZpTPaper}. 
Although the underlying dynamics being tested are similar, the uncertainties on 
the $W$ and $Z$ measurements are different and mostly uncorrelated.
The transverse energy resolution of the hadronic recoil is not as good as the resolution 
on the lepton momenta, but approximately 10 times as many candidate events are available 
($(\sigma_W\cdot BR(\Wln))/(\sigma_Z\cdot BR(\Zll)) = 10.840 \pm 0.054$~\cite{WZInclPaper}).
Testing the modeling of the hadronic recoil through the $W$ $p_T$ distribution is also an 
important input to precision measurements using the \Wln\ sample, including especially 
the $W$ mass measurement.

In this paper, we describe a measurement of the transverse
momentum distribution of $W$ bosons using ATLAS data from $pp$
collisions at \mbox{$\sqrt{s} = 7$ \TeV} at the LHC~\cite{LHCmachine}, corresponding 
to about \mbox{31~\ipb} of integrated luminosity.  The measurement is performed
in both the electron and muon channels, and the reconstructed $W$ $p_T$ 
distribution, following background subtraction, is unfolded to the true $p_T$
distribution. Throughout this paper, \ptr\ is used to refer
to the reconstructed $W$ $p_T$ and \ptw\ is used to refer to the true
$W$ $p_T$.  The true $W$ $p_T$ may be defined in three ways.  The default in this 
paper is the $p_T$ that appears in the $W$ boson propagator at the
Born level, since this definition of \ptw\ is independent
of the lepton flavor and the electron and muon measurements can be combined.
It is also possible to define \ptw\ in terms of the true lepton kinematics, 
with (``dressed'') or without (``bare'') the inclusion of QED final state
radiation (FSR).  These define a physical final state more readily identified with
the detected particles, so we give results for these definitions
of \ptw\ for the electron and muon channels.  For all three definitions of \ptw,
photons radiated by the $W$ via the $WW\gamma$ triple gauge coupling vertex are treated
identically to those radiated by a charged lepton.

The unfolding proceeds in two steps.  First, a Bayesian
technique is used to unfold the reconstructed  distribution (\ptr) to
the true distribution (\ptw) for selected events, taking into account
bin-to-bin migration effects via a response matrix describing the
probabilistic mapping  from \ptw\ to \ptr.  This step corrects for the
hadronic recoil resolution. Second, the resulting distribution is
divided in each bin by the detection efficiency, defined as the ratio
of the number of events reconstructed to the number produced in the
phase space consistent with the event selection.  This converts the
\ptw\ distribution for selected events into the \ptw\ distribution
for all $W$ events produced in the fiducial volume, which is defined by
$p_T^\ell>20$~GeV, $|\eta_\ell|<2.4$, $p_T^\nu>25$~GeV, and transverse
mass $m_T = \sqrt{2 p_T^\ell p_T^\nu (1 - \cos(\varphi^\ell - \varphi^\nu))} > 40$~GeV~\footnote{
  The origin of the ATLAS coordinate system is at the nominal $pp$ interaction point 
  in the geometrical center of the detector.  The $z$ axis points along the 
  anti-clockwise beam direction, and the azimuthal and polar angles $\varphi$ and $\eta$
  are defined in the conventional way, with $\varphi = 0$ (the $x$ axis) pointing from
  the origin to the center of the LHC ring. The pseudorapidity is defined as
  \mbox{$\eta = -\ln \tan(\theta/2)$}. The transverse momentum $p_T$, the transverse
  energy $E_T$, and the transverse missing energy \met\ are defined in the $x-y$ plane.  }, 
where the thresholds are defined in terms of the true lepton kinematics. 

The unfolding results in the differential fiducial cross section \dsdptw, 
in which the subscript in $\sigma_{\rm{fid}}$ indicates that the cross section measured
is the one for events produced within the phase space defined above.
The electron and muon differential cross sections are combined into a single
measurement via $\chi^2$ minimization, using a covariance
matrix describing all uncertainties and taking into account the correlations
between the measurement channels as well as across the \ptw\ bins.  
The resulting differential cross section is normalized to the total measured
fiducial cross section, which results in the cancellation of some uncertainties,
and compared to predictions from different
event generators and perturbative QCD (pQCD) calculations. 

This paper is organized as follows.  Section~\ref{sec:Theory} reviews the 
existing calculations and measurements of \ptw.  The relevant components 
of the ATLAS detector are described in Section~\ref{sec:Detector} and the 
generation of the simulated data used is described in Section~\ref{sec:MCSamples}.
The event selection is given in Section~\ref{sec:Selection} and the estimation
of the backgrounds remaining after that selection is explained in  
Section~\ref{sec:Backgrounds}.  The unfolding procedure is described in 
Section~\ref{sec:HadRecUnfold}.  Section~\ref{sec:Systematics} summarizes
the systematic uncertainties.  The electron and muon channel results, the 
combination procedure, and the combined 
results are all given in Section~\ref{sec:Results}.
We conclude with a discussion of the main observations in Section~\ref{sec:Conclusions}.

\section{\label{sec:Theory}QCD Predictions and Previous Measurements}

At leading order, the $W$ boson is produced with zero 
momentum transverse to the beam line.  Non-zero $p_T$ is generated through the emission 
of partons in the initial state.  At low $p_T$, this is dominated by multiple soft or
almost collinear partons, but at higher $p_T$, the emission of one or more hard partons
becomes the dominant effect.  Because of this, different calculations of
$d\sigma/d\ptw$ may be better suited for different ranges of \ptw.

At large \ptw\ ($\ptw\gsim 30$~GeV), the spectrum is determined primarily by hard parton emission, and
perturbative QCD (pQCD) calculations at a fixed order of $\alpha_s$ are expected
to predict $d\sigma/d\ptw$ reliably~\cite{Davies:1984hs}.  
The inclusive cross section prediction is finite, but the differential cross 
section diverges as \ptw\ approaches zero.  
Differential cross sections calculated to $\mathcal{O}(\alpha_s^2)$ are available for
$Z/\gamma^*$ production through the \fewz~\cite{fewz1,fewz2} and \dynnlo~\cite{dynnlo,dynnlo2} programs,
and are becoming available for the $W$.  
The \mcfm\ generator~\cite{mcfm} can predict \ptw\ at $\mathcal{O}(\alpha_s^2)$ 
through the next-to-leading order (NLO) calculation of the $W$~+~1 parton differential cross section.

As \ptw\ becomes small, contributions at higher powers of $\alpha_s$ describing 
the production of soft gluons grow in importance.  These terms also
contain factors of $\ln(M_W^2/(p_T^{W})^2)$ which diverge for vanishing \ptw.
The \ptw\ distribution is better modeled in this regime by calculations that 
resum logarithmically divergent terms to all orders in 
$\alpha_s$~\cite{Davies:1984hs,Collins:1984kg,resbos1}.  
The \resbos\ generator~\cite{resbos1,resbos4,resbos20} resums the leading contributions
up to the next-to-next-to-leading logarithms (NNLL), and matches the resummed calculation
to an $\mathcal{O}(\alpha_s)$ calculation, corrected to $\mathcal{O}(\alpha_s^2)$
using a $k$-factor depending on $p_T$ and rapidity, to extend the prediction to large \ptw.
It also includes a non-perturbative parametrization, tuned to Drell-Yan data from 
several experiments~\cite{resbos20,resbos12}, to model the lowest \ptw\ values.

Parton shower algorithms such as \pythia~\cite{pythia} and \herwig~\cite{herwig} 
can also provide finite predictions of 
$d\sigma/d\ptw$ in the low-\ptw\ region by describing the soft gluon
radiation effects through the iterative splitting and radiation of
partons. \pythia\
implements leading-order matrix element calculations with a 
parton shower algorithm that has been tuned to match the \ptz\ data from the 
Tevatron~\cite{PerugiaTunes,CdfZpT,D0ZpT2}.  Similarly, the \mcatnlo~\cite{mcatnlo} and 
\powheg~\cite{powheg1,powheg2,powheg3,powhegVB} event generators combine 
NLO ($\mathcal{O}(\alpha_s)$) matrix 
element calculations with a parton shower algorithm to produce 
differential cross section predictions that are finite for all \ptw.  

Generators such as \alpgen~\cite{alpgen} 
and \sherpa~\cite{sherpa} calculate matrix elements for higher orders in $\alpha_s$ (up to five), 
but only include the tree-level terms which describe the production of 
hard partons.  Parton shower algorithms can be run on the resulting events,
with double-counting of parton emissions in the phase space overlap between the
matrix element and parton shower algorithms removed through a veto~\cite{alpgen} 
or by reweighting~\cite{Hoeche:2009rj,ckkw}. Although these calculations do not
include virtual corrections to the LO process, they are relevant 
for comparison to the highest $p_T$ part of the \ptw\ spectrum, which includes
contributions from a $W$ recoiling against multiple high-$p_T$ jets. 

The $W$ $p_T$ distribution has been measured most recently at the Tevatron with Run I data 
($p\bar{p}$ collisions at $\sqrt{s} = 1.8 \TeV$) by both CDF~\cite{CdfRunIWpT} and 
D0~\cite{D0RunIWpT}.   Both of these results are limited by the number of candidate
events used (less than 1000), and by the partial unfolding which does not take into 
account bin-to-bin correlations. The present analysis uses more than 100,000 candidates per channel and a 
full unfolding of the hadronic recoil which takes into account correlations between bins, 
resulting in greater precision overall and inclusion of higher-\ptw\ events compared to 
the Tevatron results.

Although this is the first measurement of the $W$ $p_T$ distribution at 
the LHC, the \Wln\ sample at $\sqrt{s} = 7$~TeV has been studied recently by both the ATLAS and 
CMS collaborations.  The ATLAS collaboration has measured the inclusive \Wln\ cross section~\cite{WZInclPaper} and the lepton charge 
asymmetry in \Wmn\ events~\cite{WChargeAsymPaper}.  The CMS collaboration has also measured 
the inclusive cross section~\cite{CMSWZInclPaper}, and has measured the polarization of
$W$s produced with $\ptw > 50$~GeV, demonstrating that the majority of $W$ bosons produced 
at large $p_T$ in $pp$ collisions are left-handed, as predicted by the 
standard model~\cite{CMSWPolarization}.

\section{\label{sec:Detector}The ATLAS Detector and the $pp$ Dataset}

\subsection{The ATLAS Detector}

The ATLAS detector~\cite{DetectorPaper2008} at the LHC
consists of concentric cylindrical layers of inner tracking, calorimetry, and outer
(muon) tracking, with both the inner and outer tracking volumes contained, or
partially contained, in the fields of superconducting magnets to enable measurement 
of charged particle momenta.

The inner detector (ID) allows precision tracking of charged particles within
$|\eta| \sim 2.5$.  It surrounds the interaction point, inside
a superconducting solenoid which produces a 2~T axial field.  The
innermost layers constitute the pixel detector, arranged in three
layers, both barrel and endcap. The semi-conductor tracker (SCT) is
located at intermediate radii in the barrel and intermediate $z$ for
the endcaps, and consists of four double-sided silicon strip layers
with the strips offset by a small 
angle to allow reconstruction of three-dimensional space points.
The outer layers, the transition radiation tracker (TRT),
are straw tubes which provide up to 36 additional $R-\varphi$ position 
measurements, interleaved with thin layers of material which stimulate the 
production of transition radiation. This radiation is then detected as a higher ionization 
signal in the straw tubes, and exploited to distinguish electron from pions.

The calorimeter separates the inner detector from the muon spectrometer and measures
particle energies over the range $|\eta| < 4.9$.  The liquid argon (LAr) electromagnetic 
(EM) calorimeter uses a lead absorber in folded layers designed to minimize gaps in coverage.
It is segmented in depth to enable better particle shower reconstruction.
The innermost layer (``compartment'') is instrumented with
strips that precisely measure the shower location in $\eta$.  The middle compartment
is deep enough to contain most of the electromagnetic shower produced by a typical electron
or photon.  The outermost compartment has the coarsest spatial resolution and is used to quantify
how much of the particle shower has leaked back into the hadronic calorimeter.
The hadronic calorimeter surrounds the electromagnetic calorimeter and extends the 
instrumented depth of the calorimeter to fully contain hadronic particle showers.
Its central part, covering $|\eta| < 1.7$, is the tile calorimeter, which is constructed
of alternating layers of steel and scintillating plastic tiles.  Starting at $|\eta| \sim 1.5$
and extending to $|\eta| \sim 3.2$, the hadronic calorimeter is part of the liquid
argon calorimeter system, but with a geometry different from the EM calorimeter
and with copper and tungsten as the absorbing material. The forward
calorimeters (FCAL), also using liquid argon, extend the coverage up to $|\eta| \sim 4.9$.

The muon chambers and the superconducting air-core toroid
magnets, located beyond the calorimeters, constitute the muon spectrometer
(MS). Precision tracking in the bending plane ($R-\eta$) for both
the barrel and the endcaps is performed by means of monitored
drift tubes (MDTs).  Cathode strip chambers (CSCs) provide precision
$\eta-\varphi$ space points in the  innermost layer of the endcap, for
$2.0 < |\eta| < 2.7$.  The muon triggers are
implemented via resistive plate chambers (RPCs) and thin-gap chambers
(TGCs) in the barrel and endcap, respectively.  In addition to fast
reconstruction of three-dimensional space points for muon triggering,
these detectors provide $\varphi$ hit information complementary to
the precision $\eta$ hits from the MDTs for muon reconstruction.

\subsection{Online Selection}

The online selection of events is based on rapid reconstruction and identification of 
charged leptons, and the requirement of at least one charged lepton candidate observed 
in the event.  The trigger system implementing the online selection has three levels:
Level 1, which is implemented in hardware; Level 2, which runs specialized reconstruction
software on full-granularity detector information within a spatially limited ``Region of
Interest''; and the Event Filter, which 
reconstructs events using algorithms and object definitions nearly identical to
those used offline.  

In the electron channel, the Level 1 hardware selects events with
at least one localized region (``cluster'') of significant energy deposition in the
electromagnetic calorimeter with $E_T > 10$~GeV.  Level 2 and the Event Filter
check for electron candidates in events passing the Level 1 selection, and accept events
with at least one electron candidate with $E_T > 15$~GeV.  The electron identification includes
matching of an inner detector track to the electromagnetic cluster and requirements on 
the cluster shape.  The trigger efficiency relative to offline electrons
as defined below is close to 100\% within the statistical
uncertainties in both data and simulation. 

The online selection of muon events starts from the identification of
hit patterns consistent with a track in the muon spectrometer at Level 1.  
For the first half of the data used in this analysis, there is 
no explicit threshold for the transverse momentum at Level 1, but in the second half, to cope with 
increased rates from the higher instantaneous luminosity, a threshold of 10 GeV is used.  
Level 2 and the Event Filter attempt to reconstruct muons in events passing the Level 1 
trigger using an ID track matched to a track segment in the MS.  Both 
apply a $p_T$ threshold of 13 GeV for all of the data used in this analysis.   
The trigger efficiency relative to the offline
combined muon defined below is a function of the muon $p_T$ and $\eta$, and varies between
67\% and 96\%. Due to its larger geometrical coverage, the
endcap trigger is more efficient than the barrel trigger. The trigger
path starting from a Level 1 trigger with no explicit $p_T$ threshold
is slightly more efficient (1-2\%) than the one with a 10 GeV
threshold. 

\subsection{Data Quality Requirements and Integrated Luminosity}

Events used in this analysis were collected during stable beams operation of the LHC in 2010
at $\sqrt{s} = 7$ TeV with all needed detector components functioning nominally, 
including the inner detector, calorimeter, muon spectrometer, and magnets.  The integrated
luminosity is $31.4 \pm 1.1$~\ipb\ in the electron channel and $30.2 \pm 1.0$~\ipb\ in the muon 
channel~\cite{LumiPaper,LumiConf}.

\section{\label{sec:MCSamples}Event Simulation}

Simulated data are used to calculate the efficiency for the \Wln\ signal,
to estimate the number of background events and their distribution in \ptwreco,
to construct the response matrix, and to compare the resulting normalized
differential cross section \dsdptwNorm\ to a variety of predictions.  

The simulated \Wln\ events used to calculate the reconstruction efficiency
correction and to construct the data-driven response matrix are generated using 
\pythia\ version 6.421~\cite{pythia} with the MRST 2007 LO$^{*}$ PDF set~\cite{mrst}.  	
The electroweak backgrounds (\Wtau\ and \Zgll) are estimated using other \pythia\ 
samples generated in the same way.  Simulated \ttbar\ and single-top events are generated using
\mcatnlo\ version 3.41~\cite{mcatnlo} and the CTEQ6.6 PDF set~\cite{cteq66}.
For those samples, the \herwig\ generator version 6.510~\cite{herwig} is used for 
parton showering and \jimmy\ version 4.1~\cite{jimmy} is used to model the underlying event.
The muon channel multijet background estimate uses a set of \pythia\ dijet samples
with a generator-level filter requiring at least one muon with $|\eta| < 3.0$ and $p_T > 8$~GeV.
The multijet background estimate in the electron channel uses a
\pythia\ dijet sample with a generator-level filter requiring particles with
energy totaling at least 17~GeV in a cone of radius
$\Delta R = \sqrt{(\Delta \eta)^2 + (\Delta \varphi)^2} = 0.05$.  
In both channels, the normalization of the multijet background is set
by the data. The multijet samples are used to provide an initial estimate of the background 
in the electron channel, and to extrapolate data-driven background estimates
from control data to the signal region in the muon channel.

In all of the simulated data, QED radiation of photons from charged leptons was modeled 
using PHOTOS version 2.15.4~\cite{Photos} and taus were decayed by TAUOLA version 1.0.2~\cite{Tauola}.	
The underlying event and multiple interactions were simulated according to the ATLAS MC09
tunes~\cite{MC09tune}, which take information from the Tevatron into account.
Additional inelastic collisions so generated are overlaid
on top of the hard-scattering event to simulate the effect of multiple interactions
per bunch crossing (``pile-up'').  The number of additional interactions is randomly generated following
a Poisson distribution with a mean of two.  Simulated events are then reweighted so that the 
distribution of the number of inelastic collisions per bunch crossing matches that in 
the data, which has an average of 1.2 additional collisions.  			
The interaction of the generated particles with the ATLAS
detector was simulated by GEANT4~\cite{geant4, AtlasSimulation}.  The simulated data are
reconstructed and analyzed with the same software as the $pp$ collision data.

The electroweak and top quark background predictions are normalized using the calculated
production cross sections for those processes.  For $W$ and $Z$ backgrounds, the cross
sections are calculated to next-to-next-to-leading-order (NNLO) using
FEWZ~\cite{fewz1,fewz2} with the MSTW 2008~\cite{mstw2008} PDFs (see Ref.~\cite{WZInclPaper} for details).  
The $\ttbar$ cross section is calculated at NLO with the leading NNLO terms 
included~\cite{MochUwerTop}, setting 
$m_t = 172.5$ GeV and using the CTEQ6.6 PDF set.  The single-top cross section 
is calculated using \mcatnlo\ with $m_t = 172.5$ GeV and using the CTEQ6M PDF set.

We correct simulated events for differences with
respect to the data in the lepton reconstruction and identification
efficiencies as well as in energy (momentum) scale and resolution. The
efficiencies are determined from selected $W$ and $Z$ events, using
the ``tag-and-probe'' method~\cite{WZInclPaper}. The resolution and scale
corrections are obtained from a fit to the observed $Z$ boson line shape. 

Additional \Wln\ samples from event generators other than \pythia\ are used for comparison
with the measured differential cross section \dsdptwNorm.  The \mcatnlo\ sample used is generated
with the same parameters as the \ttbar\ sample described above.
The \powheg\ events are generated using the same CTEQ6.6 PDF 	
set as the main \pythia\ \Wln\ samples, and \powheg\ is interfaced to \pythia\ for parton showering 
and hadronization.  \alpgen\ version 2.13~\cite{alpgen} matrix element calculations are interfaced to 
the \herwig\ version 6.510~\cite{herwig} parton shower algorithm,   
and use \jimmy\ version 4.31~\cite{jimmy} to model the underlying event contributions. 
These events are generated using the CTEQ6L1 PDF set~\cite{cteq6L1}. 
\sherpa\ event generation was done using version 1.3.0~\cite{sherpa}, which includes 
a Catani-Seymour subtraction based parton shower model~\cite{Schumann:2007mg}, 
matrix element merging with truncated showers~\cite{Hoeche:2009rj} and 
high-multiplicity matrix elements generated by {\sc Comix}~\cite{Gleisberg:2008fv}.
The CTEQ6L1 PDF set is used, and the renormalization and factorization scales are set dynamically
for each event according to the default \sherpa\ prescription.

\section{\label{sec:Selection}Reconstruction and Event Selection}

The \ptw\ measurement is performed on a sample of candidate \Wln\ events, 
which are reconstructed in the final state with one high-$p_T$ electron or muon and 
missing transverse energy sufficient to indicate the presence of a neutrino.
    
The event selection used in this paper closely follows that used 
in the inclusive $W$ cross section measurement presented in Ref.~\cite{WZInclPaper}.  
The selection in the muon channel is identical to that used in the $W$ lepton charge asymmetry 
measurement in Ref.~\cite{WChargeAsymPaper}.  The event reconstruction and $W$ candidate 
selection are summarized here.

\subsection{Lepton ($e$, $\mu$, and $\nu$) Reconstruction}

Electrons are reconstructed as inner detector tracks pointing to particle showers reconstructed
as a cluster of cells with significant energy deposition in the electromagnetic calorimeter.  
This analysis uses electrons with clusters fully contained in either the barrel
or endcap LAr calorimeter. These requirements translate into $|\eta_e| < 2.47$ with the transition region 
$1.37 < |\eta_e| < 1.52$ excluded.  To reject background (essentially
originating from hadrons), multiple requirements 
on track quality and the electromagnetic shower profile are applied, following the 
``tight'' selection outlined in Ref.~\cite{WZInclPaper}.  
Track quality criteria include a minimum number of hits in the pixel detector, SCT, and TRT,
as well as requirements on the transverse impact parameter and a 
minimum number of TRT hits compatible with the
detection of X-rays generated by the transition radiation from
electrons.
The energy deposition pattern in the calorimeter is characterized by its depth as well as
its width in the three compartments of the LAr calorimeter, and the parameters are
compared with the expectation for electrons. The position of the reconstructed cluster is 
required to be consistent with the location at which the extrapolated electron track
crosses the most finely-segmented part of the calorimeter. Since electron
showers are expected to be well contained within the LAr calorimeter, electron
candidates with significant associated energy deposits in the tile
calorimeter are discarded. Finally, electron candidates compatible with
photon conversions are rejected.  
Although there is no explicit isolation requirement in the electron identification for this 
analysis, the criteria selecting a narrow shower shape in the calorimeter
provide rejection against non-isolated electrons from heavy flavor decays. With these
definitions, the average electron selection efficiency ranges from
67\% in the endcap ($1.52 < |\eta| < 2.47$) to 84\% in the central
region ($|\eta|<1.37$) for simulated $W$ events.

Muons are reconstructed from tracks in the muon spectrometer joined to tracks in 
the inner detector.  The track parameters of the combined muon are the statistical 
combination of the parameters of the MS and ID tracks, where the track parameters are
weighted using their uncertainties for the combination.  Combined muon candidates with 
$|\eta| < 2.4$, corresponding to the coverage of the RPC and TGC detectors used in
the trigger, are used in this analysis.  To reject backgrounds from meson 
decays-in-flight and other poorly-reconstructed tracks, the $p_T$ measured using the
MS only must be greater than 10 GeV, and the $p_T$ measured in the MS and ID must be 
kinematically consistent with each other: 
\begin{equation}
|p_T^{\mathrm{MS}} ({\mathrm{energy\;loss\;corrected}})- p_T^{ID}|<0.5\phantom{0}p_T^{\mathrm{ID}}.
\end{equation}
For both of these requirements, the momentum measured in the muon spectrometer is corrected 
for the ionization energy lost by the muon as it passes through the calorimeter.  There 
are no explicit requirements on the number of hits associated with the MS track, but the ID 
track is required to have hits in the pixel detector, the SCT, and the TRT, although if the track
is outside of the TRT acceptance that requirement is omitted.  Finally, to reject background
from muons associated with hadronic activity, particularly those produced by the decay of a 
hadron containing a bottom or charm quark, the muon is required to be isolated.  The
isolation is defined as the scalar sum of the $p_T$ of the ID tracks immediately surrounding 
the muon candidate track ($\Delta R < 0.4$).  The isolation threshold scales with the muon 
candidate $p_T$ and is $\sum p_T^{\mathrm{ID}} < 0.2\,p_T^\mu$.  The combined muon 
reconstruction and selection efficiency varies from 90\% to 87\% as the muon $p_T$ increases
from 20~GeV to above 80~GeV.

The transverse momentum of the neutrino produced by the $W$ decay can be approximately 
reconstructed via the transverse momentum imbalance measured in the detector, also known 
as the missing transverse energy (\met).  The \met\ calculation begins from the negative of 
the vector sum over the whole detector of the momenta of clusters in 
the calorimeter.  The magnitude and position of the energy deposition determines the momentum
of the cluster.  The cluster energy is initially measured at the electromagnetic scale, under the 
assumption that the only energy deposition mechanism is electromagnetic showers such as those produced
by electrons and photons.   The cluster energies are then corrected 
for the different response of the calorimeter to hadrons relative to electrons and photons,
for losses due to dead material, and for energy which is not captured by the clustering process.
The \met\ used in the electron channel is exactly this calorimeter-based calculation.
In the muon channel, the \met\ is additionally corrected for the fact that muons, 	
as minimum ionizing particles, typically only lose a fraction of their momentum in the calorimeter.
For isolated muons, the \met\ is
corrected by adding the muon momentum as measured with the combined ID and MS track to the
calorimeter sum, with the calorimeter clusters associated with the energy deposition of the 
muon subtracted to avoid double-counting.
In this context, muons are considered isolated if the $\Delta R$ to the nearest jet with $E_T > 7$~GeV
is greater than 0.3.  Jets are reconstructed using the anti-$k_t$ algorithm~\cite{antikt}
and the $E_T$ is measured at the electromagnetic scale.
For non-isolated muons, the muon momentum is measured using only the muon
spectrometer. In this case, the momentum loss in the calorimeter is
kept within the calorimeter sum. To summarize, the \met\ is calculated
via the formula  
\begin{equation}
E^{\rm{miss}}_{x,y} \ = \   -  \sum_{i} E_{x,y}^i
\ - \ \sum^{\rm isolated}_j p_{x,y}^j  \  -  \  \sum^{\rm non-isolated}_k p_{x,y}^k \;.
\label{eq:met}
\end{equation}
In the above, the $E_{x,y}^i$ are the individual topological cluster momentum components,
excluding those clusters associated with any isolated muon,
the $p_{x,y}^j$ are the momenta of isolated muons as measured with the combined track,
and the $p_{x,y}^k$ are the momenta of non-isolated muon as measured in the muon spectrometer.
In practice, for the electron channel, only the first 
term contributes, but for the muon channel all three terms contribute.

\subsection{Event Selection}

Candidate $W$ events are selected from the set of events passing a single electron or a 
single muon trigger.  Offline, events are first subject to cleaning requirements
aimed at rejecting events with background from cosmic rays or detector noise. 
These requirements reject a small fraction of the data  
and are highly efficient for the $W$ signal~\cite{WZInclPaper}.
Events must have a reconstructed primary vertex with at least three tracks with $p_T > 150$~MeV.  They 
are rejected if they contain a jet with features characteristic of a known non-collision 
localized source of apparent energy deposition, such as electronic noise in the calorimeter.
Such spurious jets can result in events with large \met\ but which do not contain
a neutrino or even necessarily originate from a $pp$ collision.
In the electron channel, events are rejected if the electron candidate is reconstructed
in a region of the calorimeter suffering readout problems during the 2010
run~\cite{CaloCleaning}. This last requirement results in 
a~$\sim 5\%$ efficiency loss. 

After the event cleaning, we select events with at least one electron or 
muon, as defined above, with transverse momentum greater than 20
GeV. In events with more than one such lepton, the lepton with largest
transverse momentum is assumed to originate from the $W$ decay. To
provide additional rejection of cosmic rays, muon candidates must
point at the primary vertex, in the sense that the offset in $z$ along
the beam direction between the primary vertex and the   point where
the candidate muon track crosses the beam line must be less than
10~mm. 

Finally, we require $\met > 25$~GeV and transverse mass greater than 40 GeV
to ensure consistency of the candidate sample with the expected kinematics of $W$ decay.

After all selections, 112 909 \Wen\ candidates and 129 218
\Wmn\ candidates remain in the data.  The smaller number of candidate events
in the electron channel is mostly due the lower electron
reconstruction and identification efficiency.

\subsection{\label{hadrecdef}Hadronic Recoil Calculation}

The reconstruction of the $W$ boson transverse momentum is based on a
slight modification of the \met\ calculation described above.
Formally, the $\vec{p}_T$ of the $W$ boson is reconstructed as the
vector sum of the $\vec{p}_T$ of the neutrino and the charged lepton,
$\ptwvec = \vec{p}_T^\ell + \vec{p}_T^\nu$.  But the neutrino $p_T$ is
reconstructed  through the \met, and the \met\ is determined in part
from the lepton momentum, explicitly in the case  of \Wmn\ events, and
implicitly in \Wen\ events through the sum over calorimeter
clusters. Therefore when the $\vec{p}_T$ of the charged lepton and
\met\ are summed, the charged lepton momentum cancels out and the
$W$ transverse momentum is measured as the summed $\vec{p}_T$ of the  calorimeter
clusters, excluding those associated with the electron or muon. This
part, which consists of the energy deposition of jets and softer
particles not clustered into jets, is referred to as the hadronic
recoil \hr. The reconstructed \ptw\ is denoted \ptr\ and is defined as
the magnitude of \hr.

In this measurement, the exclusion of the lepton from \ptr\ is made explicit 
by removing all clusters with a $\Delta R < 0.2$ relative to the charged lepton.
This procedure leaves no significant lepton flavor dependence in the
reconstruction of \ptr, so that it is possible to construct a combined
response matrix describing the mapping from \ptw\ to \ptr\ which can
be applied to both channels. To compensate for the energy from
additional low-$p_T$ particles removed along with the lepton, the
underlying event is sampled on an event-by-event basis using a cone of
the same size, placed at the same $\eta$ as the lepton. The cone
azimuth is randomly chosen but required to be away from the lepton and
original recoil directions, to ensure that the compensating energy is
not affected significantly by these components of the event. The
distance in azimuth to the lepton is required to satisfy
$\Delta\phi>2\times\Delta R$, and the distance to the recoil should
match $\Delta\phi>\pi/3$. The transverse momentum measured from
calorimeter clusters in this cone is rotated to the position of the removed lepton
and added to the original recoil estimate. Because this procedure is
repeated for every event, the energy in the clusters in the
replacement cone contains an amount of energy from the underlying
event and from multiple proton-proton collisions (``pile-up'') which
is correct on average for each event and accounts for event-by-event
fluctuations. 

\section{\label{sec:Backgrounds}Background Estimation} 

\begin{figure*}[tbp]
    \centering
    \begin{overpic}[width=0.49\textwidth]{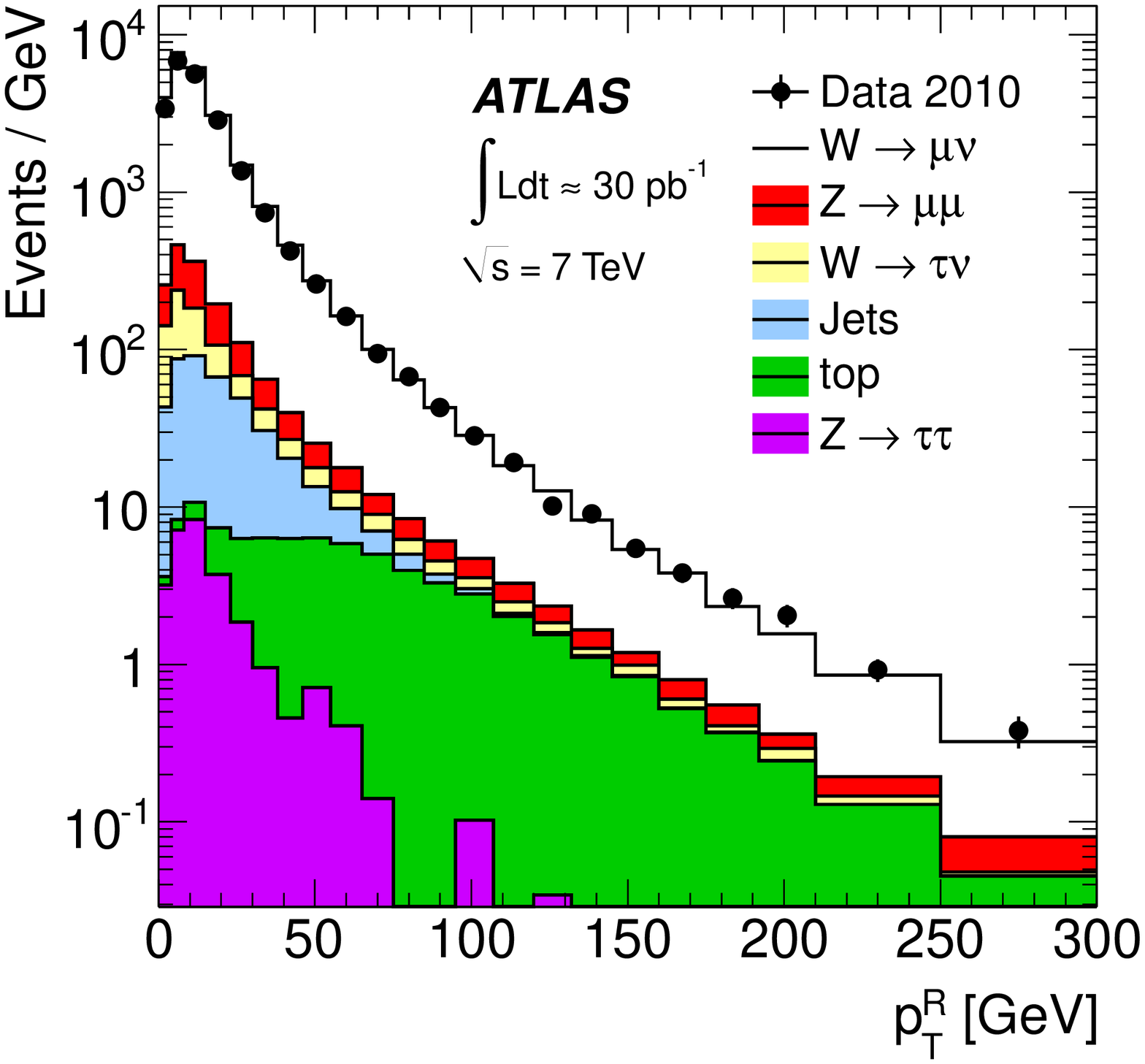}
        \put(30,82){(a)}
    \end{overpic}
    \begin{overpic}[width=0.49\textwidth]{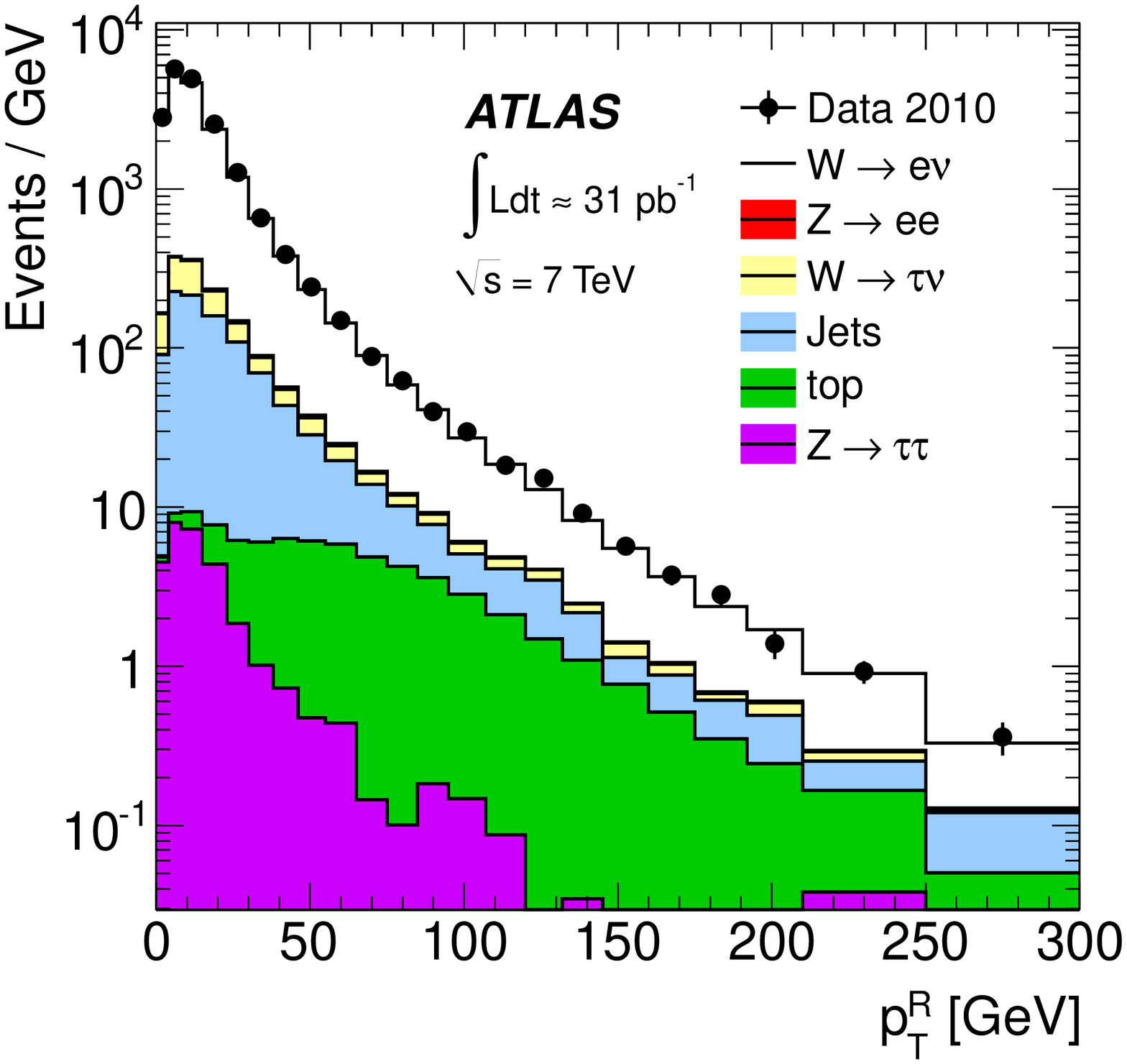}
        \put(30,82){(b)}
    \end{overpic}
\caption{\label{fig:ptwBackgrounds}
    Observed and predicted \ptr\ distributions in the electron
    channel (a) and in the muon channel (b).}
\end{figure*}

Backgrounds to \Wen\ and \Wmn\ events come from other types of electroweak
events (\Zll\ and \Wtau), \ttbar\ and single top events, and from multijet events
in which a non-prompt lepton is either produced through the decay of a hadron
containing a heavy quark ($b$ or $c$), the decay-in-flight of a light meson to a 
muon, or through a coincidence of hadronic signatures that mimics the characteristics of a lepton.
Figure~\ref{fig:ptwBackgrounds} shows the expected and
observed \ptr\ distribution in the electron and muon channels,
with background contributions calculated as described below.

Electroweak backgrounds (\Wtau, \Zll, \Ztau) and top quark production 
(\ttbar\ and single top) are estimated using the acceptance and efficiency
calculated from simulated data, corrected for the imperfect detector simulation
and normalized using the predicted cross sections as described in Section~\ref{sec:MCSamples}. 
These backgrounds amount to about 6\% of the selected events in the electron channel, and to
about 10\% in the muon channel.  The background in the muon channel is larger
because the smaller geometrical acceptance of the muon spectrometer compared
to the calorimeter leads to a greater contribution of \Zmm\ events compared 
to \Zee\ events.
Uncertainties on the summed electroweak and 
top background rates are 6\% at low \ptr\ in both channels, rising to 14\% above
$\ptr\sim 200$~GeV in the muon channel, and 25\% in the electron channel. 
The leading uncertainties on these backgrounds at low \ptr\ are from the theoretical
model, since the cross sections used to normalize them have uncertainties
ranging from 4\% (for $W$ and $Z$) to 6\% (for \ttbar), and from the
PDF uncertainty on the acceptances, which is 3\%~\cite{WZInclPaper}.  The
integrated luminosity calibration contributes an additional 3.4\%~\cite{LumiPaper,LumiConf}.  
Important experimental uncertainties include the 
energy (momentum) scale uncertainty, which contributes about 3\% (1\%) at low 
\ptr\ in the electron (muon) channel, increasing to about 6\% (5\%) at high \ptr.  
At high \ptr\ ($\ptr \gtrsim 150$~GeV), there are also significant contributions
for both channels from the statistical uncertainty on the acceptance and efficiency 
calculated from simulated events.

The multijet backgrounds are determined using data-driven methods. In the
electron channel, the observed \met\ distribution is interpreted in
terms of signal and background contributions, using a method based on
template fitting. A first template is built from the signal as well as
electroweak and top backgrounds, using simulated events. The
multijet background template is built from a background-enriched
sample, obtained by applying all event selection cuts apart  
from inverting a subset of the electron identification criteria. The
multijet background fraction is then determined by a fitting procedure
that adjusts the normalization of the templates to obtain the best
match to the observed \met\ distribution. This method has been
described in Ref.~\cite{WZInclPaper}, and is applied here bin by bin
in \ptr. The multijet background fraction is 4\% at low \ptr, and
rises to $\sim 9$\% at high \ptr. Uncertainties on this method are
estimated from the stability of the fit result under different event
selections used to produce the multijet background templates, by
propagating the lepton efficiency and momentum scale uncertainties to
the signal templates, and by varying the range of the \met\
distribution used for the fit. These sources amount to a total
relative uncertainty of 25\% at low \ptr, decrease to 5\% at $\ptr\sim
35$~GeV, and progressively rise again to 100\% at high \ptr, where
very few events are available to construct the templates.

In the muon channel, the multijet background is primarily from semi-leptonic heavy
quark decays, although there is also a small component from kaon or pion 
decays-in-flight. The estimation of this background component relies on
the different efficiencies of the isolation requirement for multijet and
electroweak events, and is based on the method described in Ref.~\cite{WZInclPaper}.  
Muons from electroweak boson decays, including those from top quark decays, 
are mostly isolated, and their isolation efficiency is measured from
$Z \rightarrow \mu\mu$ events. The efficiency of the isolation requirement
on multijet events is measured using a background-enriched control sample,
which consists of events satisfying all of the signal event selection except
that the muon transverse momentum range is restricted to $15<p_T^\mu<20$~GeV
and the \met\ and $m_T$ requirements are dropped. The
measured efficiency is extrapolated to the signal region 
($p_T^\mu>20$~GeV, $\met>25$~GeV, and $m_T>40$~GeV) using simulated multijet events. 
Knowledge of the isolation efficiency for both components, combined with
the number of events in the \Wmn\ candidate sample before and after the 
isolation requirement, allows the extraction of the multijet background. 
As for the electron channel, this method is
applied for each bin in \ptr, with the number of total and isolated candidates,
as well as the signal and background efficiencies, calculated separately for each bin. 
The isolation efficiency for the background is fitted with an exponential distribution
to smooth out statistical fluctuations arising from the limited number of 
events passing all of the event selection in the simulated multijet data.

The multijet background fraction in the muon channel is found to
be 1.5\% at low \ptr\ and decreases to become negligible for
$\ptr>100$~GeV. Uncertainties on the estimated multijet background include 
all statistical uncertainties, including those on both the signal and 
background isolation efficiency measurements.  The full range of the 
simulation-based extrapolation of the isolation efficiency for the multijet
background is taken as a systematic uncertainty.  Subtraction of residual 
electroweak events in the control samples is also included in the systematic
uncertainty but is a sub-dominant contribution.
The relative uncertainty on the background rate varies between 25\%
and 80\%, with the largest uncertainties for $\ptr<40$~GeV.

\section{\label{sec:HadRecUnfold}Unfolding of the \ptr\ Distribution}

The unfolding of the \ptr\ distribution to the \ptw\ distribution is performed in two steps.  
In the first step, the 
background-subtracted \ptr\ distribution is unfolded to the true \ptw\ distribution, using
the response matrix to model the migration of events among bins caused by the finite resolution
of the detector.  The result of this step is the distribution of $dN/d\ptw$ of all 
reconstructed $W$ events.
In the second step, this distribution is divided by a  
reconstruction efficiency correction relating the number of reconstructed $W$ events to the number of
generated fiducial $W$ events within each bin.  That correction results in the differential
cross section \dsdptw.

\subsection{\label{sec:Unfolding}Unfolding of the Recoil Distribution}

The response matrix describes the relation between \ptw\ and \ptr, the true 
and reconstructed $W$ $p_T$, respectively.  It reflects the physics 
of the process (hadronic activity from soft and hard QCD interactions) as
well as the response of the calorimeters to low energy particles. 
This is in principle captured by a response matrix drawn from simulated
\Wln\ events, but the simulation of both aspects carries significant uncertainty.  
Therefore, the treatment of the response matrix includes corrections from $Z$ data 
to improve the model.

The \Zee\ and \Zmm\ data are used
as a model for the hadronic recoil response in $W$ events because the underlying physics is 
similar but there are two independent ways to measure the $p_T$ of the $Z$, 
through the hadronic recoil or the $p_T$ of the charged leptons.  The 
lepton energy resolution is sufficiently good that
the dilepton $p_T$ can be used to calibrate the hadronic recoil, with the
dilepton $p_T$ standing in for the true $p_T$ and the hadronic recoil remaining
the ``measured'' quantity.  One could construct a response matrix purely
from \Zll\ events, but such a matrix would be limited by the relatively small 
number of \Zll\ events in the 2010 data and residual differences between $W$ and
$Z$ kinematics and production mechanisms.  To incorporate the best features of
both the $W$ simulation  and $Z$ data models,
we introduce a parametrization of the hadronic recoil scale and resolution.  
Fits to the real and simulated $Z$ data using this parametrization are used to correct
the simulated $W$ response, and the resulting corrected parametrization is used
to fill the response matrix used for the unfolding.

Following this logic, the response matrix is built in three steps.
A first version of the response matrix, denoted $M_{\rm{MC}}$, is directly 
filled from simulated \Wln\ signal events as the two-dimensional 
distribution of \ptr\ and \ptw.
The parametrized response matrix $M_{\rm{param}}$ is also based solely 
on simulated \Wln\ events but is constructed from a fit to the recoil
as described below.  The final corrected parametrized
response matrix $M_{\rm{param}}^{\rm{corr}}$ uses the same functional 
form as $M_{\rm{param}}$, but with the fit parameters corrected using
the response measured in \Zll\ data.  Only $M_{\rm{param}}^{\rm{corr}}$ 
is used in the central value of the measurement, but $M_{\rm{MC}}$ and
$M_{\rm{param}}$ are used in assessing systematic uncertainties, particularly
those arising from the response matrix parametrization and the unfolding
procedure.

To facilitate the incorporation of corrections from the $Z$ data, we introduce an 
analytical representation of the detector response to \ptw, and approximate
$M_{\rm{MC}}$ via a smearing procedure. Decomposing \hr\ into its
components parallel and perpendicular to the $W$ line of flight, \hrpara\ and   
\hrperp, the response is observed to behave as a 	
Gaussian distribution with parameters governed by \ptw\ and \sumet,
where \sumet\ is the scalar sum of the transverse energy of all calorimeter 
clusters in the event.  By choosing the coordinate system to align with the
$W$ line of flight, any scale offset (``bias'') is in the \hrpara\ direction by construction, 
and the Gaussian resolution function is centered at zero in the \hrperp\ direction.
Specifically, the approximated response $M_{\rm{param}}$ is obtained from the Monte
Carlo signal sample as follows: 
\begin{eqnarray}
\hrpara( \ptw, \sumet ) &=& \ptw + G[ \,\,\, b(\ptw), \,\,\,
  \sigma_{\parallel}( \ptw, \sumet) \,\,\, ],\label{eq:hr1}\\
\hrperp( \ptw, \sumet ) &=& G[\,\,\, 0, \,\,\, \sigma_{\perp}(
  \ptw, \sumet) \,\,\, ].\label{eq:hr2}
\end{eqnarray}
\noindent where $G$ denotes a Gaussian random number, and its
parameters $b$, $\sigma_{\parallel}$ and $\sigma_{\perp}$ are
the Gaussian mean and resolution parameters determined from fits to the
simulation.  The bias is described according to $b(\ptw) = b_0 +
b_1 \sqrt{\ptw}$, independently of \sumet.  The resolutions follow
$\sigma_{\parallel}(\ptw,\sumet) = \sigma_{\parallel,0}(\ptw) + \sigma_{\parallel,1}(\ptw) \times \sqrt{\sumet}$ 
and 
$\sigma_{\perp}(\ptw,\sumet) = \sigma_{\perp,0}(\ptw) + \sigma_{\perp,1}(\ptw) \times \sqrt{\sumet}$, 
where the \ptw\ dependence indicates that the fit is performed separately in three regions
of \ptw\ ($\ptw<8$~GeV, $8<\ptw<23$~GeV, and $\ptw>23$~GeV).  
The separation of the fit into regions of \ptw\ improves the quality of the fit.  

With the parametrization defined, it is possible to build up a response
matrix from a set of events using a smearing procedure.  Given the \ptw\ 
and \sumet\ of each event, \hrpara\ and \hrperp\ can be constructed using
random numbers distributed according to Equations~\ref{eq:hr1} and~\ref{eq:hr2}.
Then \ptr\ is reconstructed from \hrpara\ and \hrperp, and 
the results are used to fill the relationship between \ptw\ and \ptr.
Applying this procedure to the simulated signal sample results 
in the approximate response matrix $M_{\rm{param}}$.

Corrections to this parametrization are derived from \Zll\ events by
applying the same procedure to both real and simulated $Z$ events and using
the measured decay lepton pair momentum \ptll\ as the estimator of the
true $Z$ boson transverse momentum.  The hadronic recoil calculated as described
in Section~\ref{hadrecdef} has no dependence on the lepton flavor, and 
consistent response is observed in \Zee\ and \Zmm\ events.  Therefore we fit
the combined data from both channels to minimize the statistical uncertainty.
The corrected smearing parameters are defined as follows:
\begin{eqnarray}
b^{W,\rm{corr}} &=& b^{W,\rm{MC}} + (b^{\ell\ell,\rm{data}} - b^{\ell\ell,\rm{MC}} ),\\
\sigma_{\parallel}^{W,\rm{corr}} &=& \sigma_{\parallel}^{W,\rm{MC}} + (\sigma_{\parallel}^{\ell\ell,\rm{data}} - \sigma_{\parallel}^{\ell\ell,\rm{MC}} ), \textrm{ and} \\
\sigma_{\perp}^{W,\rm{corr}} &=& \sigma_{\perp}^{W,\rm{MC}} + (\sigma_{\perp}^{\ell\ell,\rm{data}} - \sigma_{\perp}^{\ell\ell,\rm{MC}} ).
\end{eqnarray}
\noindent Above, $b^{\ell\ell,\rm{data}}$ and $b^{\ell\ell,\rm{MC}}$ are determined as a
function of \ptll, and then used as a function of \ptw; $b^{W,\rm{corr}}$ 
and $b^{W,\rm{MC}}$ are functions of \ptw\ throughout. All resolution
parameters are functions of the reconstruction-level \sumet. This defines the
final, corrected response matrix $M_{\rm{param}}^{\rm{corr}}$ used in the
hadronic recoil unfolding.

\begin{figure*}[tbp]
  \begin{overpic}[width=0.49\textwidth]{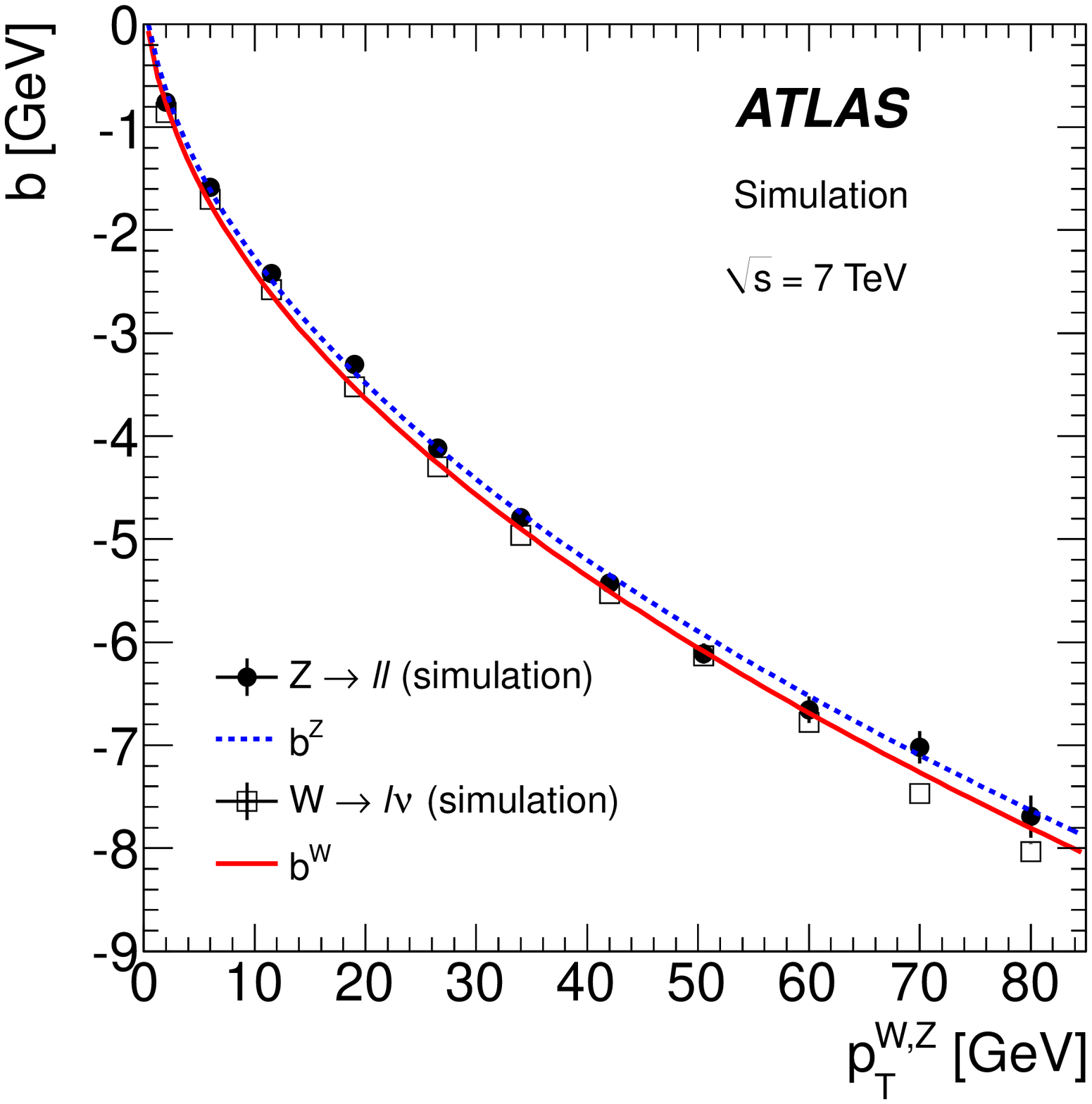}
    \put(25,82){(a)}
  \end{overpic}
  \begin{overpic}[width=0.49\textwidth]{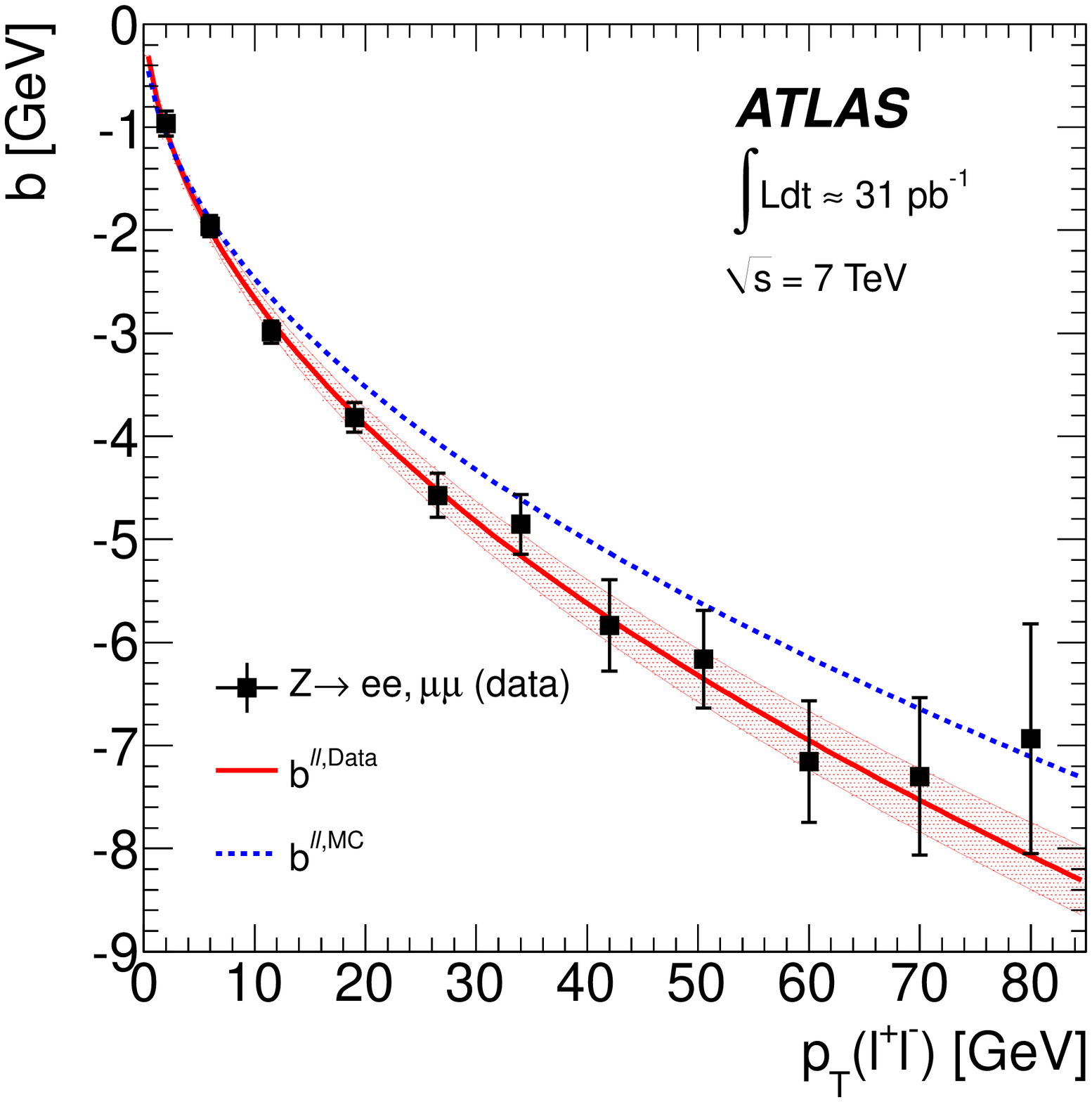}
     \put(25,82){(b)}
  \end{overpic}
  \caption{\label{fig:mcbias} {\bf (a)} Parametrization of the recoil bias as a function of the vector boson transverse momentum, $b(p_T^{W,Z})$, in
    $W$ simulation (open squares, solid line) and $Z$ simulation (solid
    circles, dashed line). {\bf (b)} Parametrization of the recoil bias as a function of the reconstructed lepton pair transverse momentum, $b(\ptll)$, in $Z$
    simulation (dashed line) and data (solid squares, shaded band).
  The shaded band shows the uncertainty on the fit.}
\end{figure*}

\begin{figure*}[htbp]
  \begin{overpic}[width=0.49\textwidth]{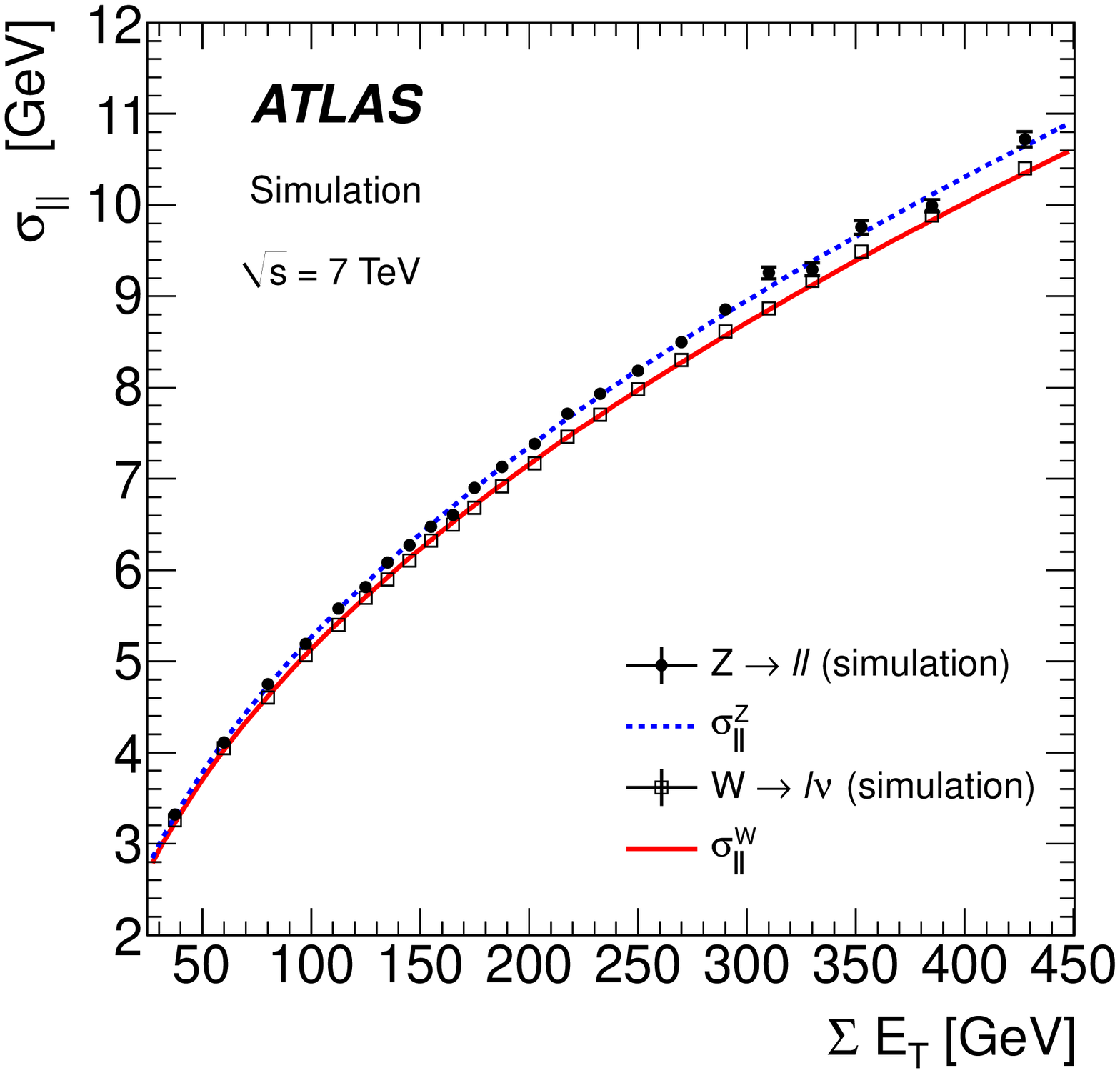}
    \put(27,23){(a)}
  \end{overpic}
  \begin{overpic}[width=0.49\textwidth]{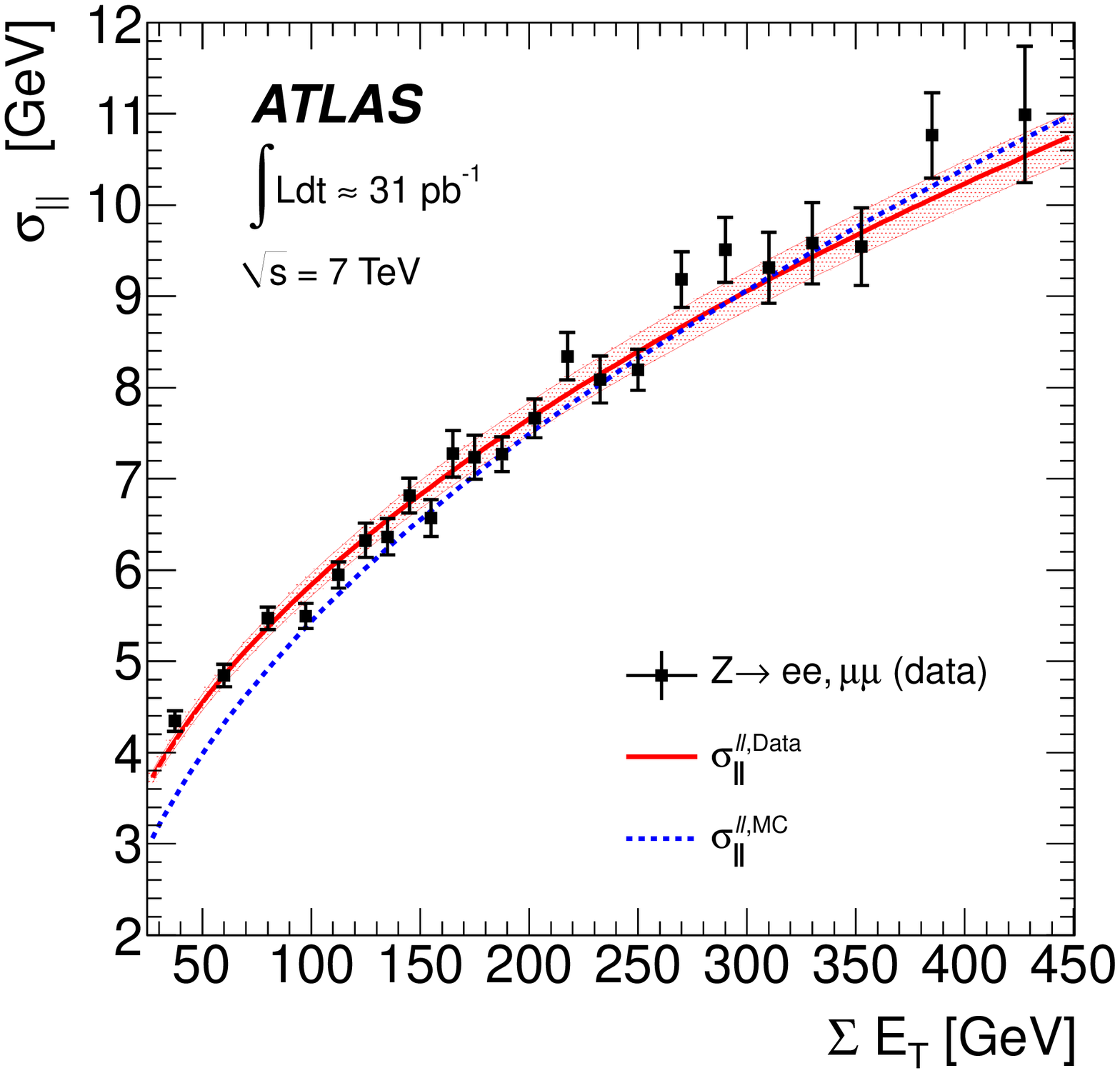}
    \put(27,23){(b)}
  \end{overpic}
  \caption{\label{fig:mcresolpara} {\bf (a)} Parametrization of the recoil resolution $\sigma_{\parallel}(\sumet)$ in
    $W$ simulation (open squares, solid line) and $Z$ simulation (solid
    circles, dashed line). {\bf (b)} Parametrization of the recoil resolution $\sigma_{\parallel}(\sumet)$ in $Z$
    simulation (dashed line) and data (solid squares, shaded band).
  The shaded band shows the uncertainty on the fit.}
\end{figure*}

\begin{figure*}[htbp]
  \begin{overpic}[width=0.49\textwidth]{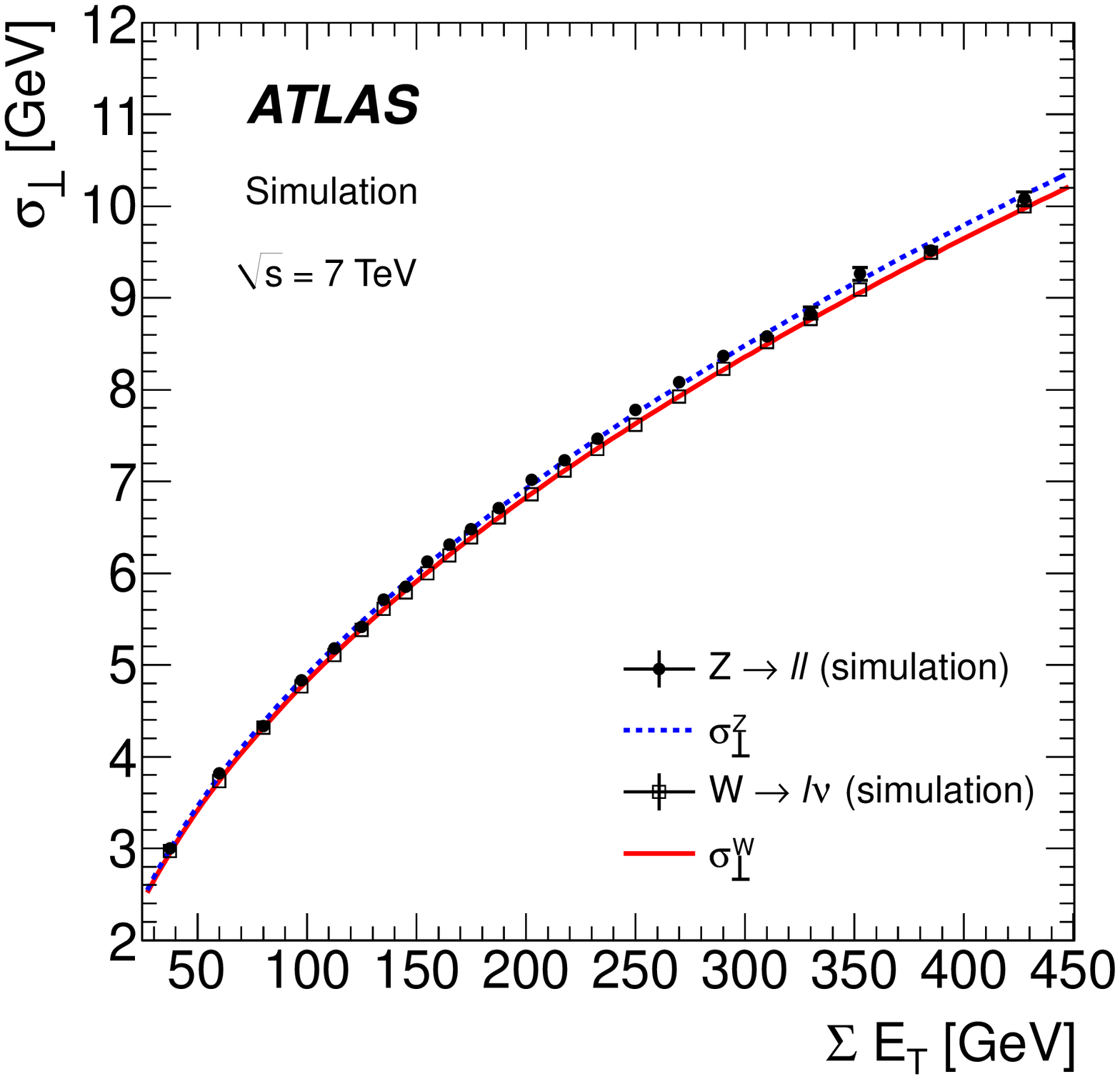}
    \put(27,23){(a)}
  \end{overpic}
  \begin{overpic}[width=0.49\textwidth]{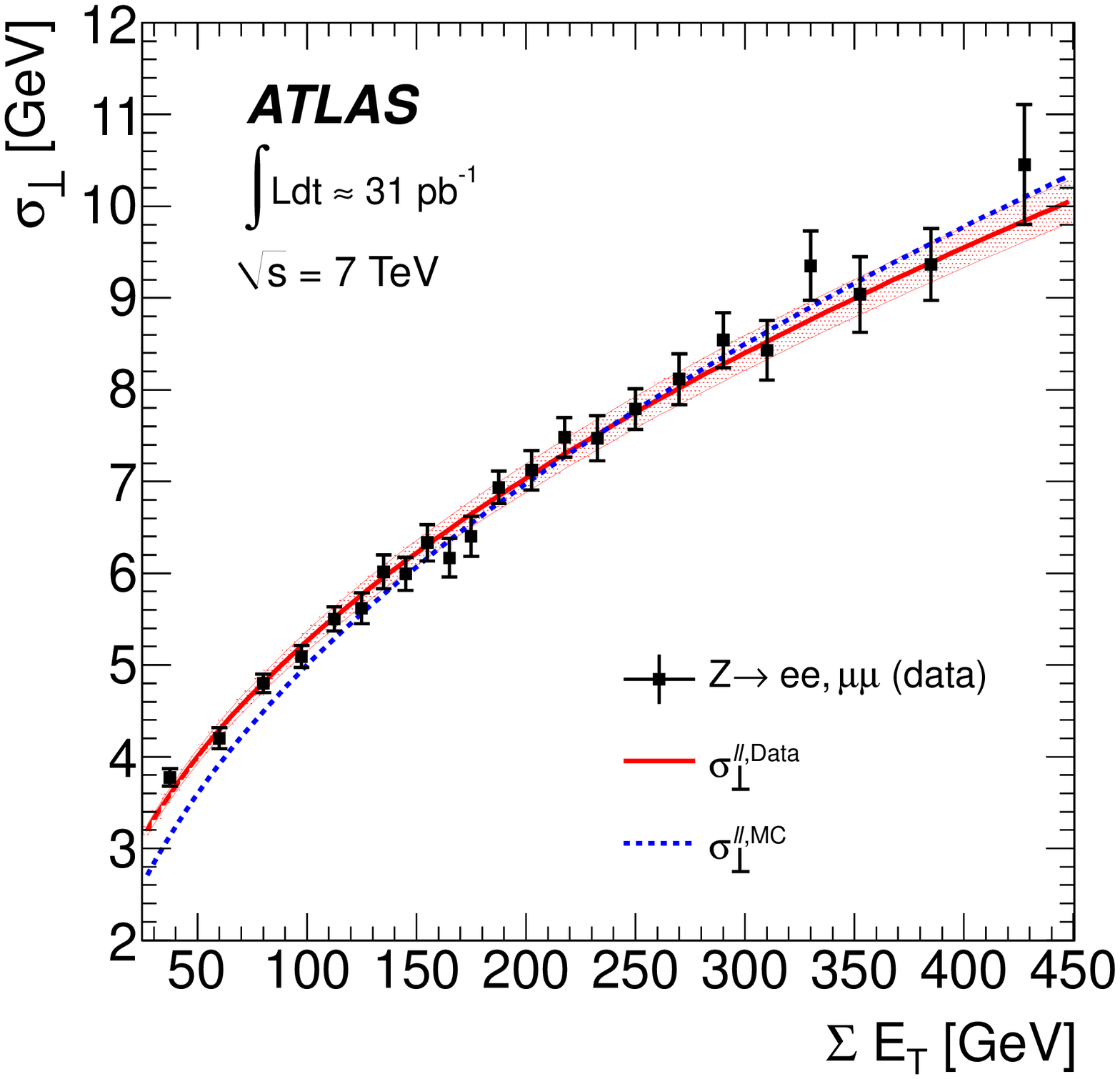}
    \put(27,23){(b)}
  \end{overpic}
  \caption{\label{fig:mcresolperp} {\bf (a)} Parametrization of the recoil resolution $\sigma_{\perp}(\sumet)$ in
    $W$ simulation (open squares, solid line) and $Z$ simulation (solid
    circles, dashed line). {\bf (b)} Parametrization of the recoil resolution $\sigma_{\perp}(\sumet)$ in $Z$
    simulation (dashed line) and data (solid squares, shaded band).
  The shaded band shows the uncertainty on the fit.}
\end{figure*}

The parametrization of the bias and resolution parameters in $W$ and
$Z$ simulation are illustrated in
Figures~\ref{fig:mcbias}(a)-\ref{fig:mcresolperp}(a). 
For these, the bias and resolution are defined with respect 
to the true (propagator) $W$ and $Z$ momenta.
The simulated and data-driven bias and resolution parameters in $Z$ events are displayed
in Figs.~\ref{fig:mcbias}(b)-\ref{fig:mcresolperp}(b). 
For these, the bias and resolution are defined with
respect to the reconstructed dilepton $p_T$. 
In Figs.~\ref{fig:mcbias}(a) and \ref{fig:mcbias}(b), the bias parametrization is shown
only over the range which determines the fit parameters,
but the parametrization describes the data well up to \ptw=300~GeV.

The response matrix is constructed using the following bin edges, expressed in~GeV: 
\begin{itemize}
\item{Reconstruction-level distribution: 0, 4, 8, 15, 23, 30, 38, 46, 55, 65, 75, 85, 95, 107, 120, 132, 145, 160, 175, 192, 210, 250, 300.}
\item{Unfolded distribution: 0, 8, 23, 38, 55, 75, 95, 120, 145, 175, 210, 300.}
\end{itemize}
The reconstruction-level binning enables more detailed comparisons
between data and simulation before unfolding, and allows a more precise 
background subtraction as a function of \ptr. It has been used in
Figure~\ref{fig:ptwBackgrounds}. The bin edges at the unfolded level 
provide a purity of at least 65\% across the \ptw\ spectrum, which 
is large enough to ensure the stability of the unfolding procedure. 
The bins are still small enough to keep the model dependence of the 
result, which enters through the assumption of a particular \ptw\ shape
within each bin, to a sub-leading contribution to
the overall uncertainty (see the description of the systematic uncertainties
in Section~\ref{sec:Systematics}). The purity is defined as the fraction of events 
where the event falls in the same bin when the bin edges are defined using 
\ptr\ as it does when the bin edges are defined using \ptw.

The unfolding of the hadronic recoil is performed by means of the
iterative Bayesian algorithm~\cite{iterDAgostini}, where the \ptw\ distribution
predicted by the simulation is used as first assumption of the true
\ptw\ spectrum, and iteratively updated using the observed
distribution. This procedure converges after three iterations. 

The statistical uncertainty on the unfolded spectrum is obtained by
generating random replicas of the reconstruction-level data.  First, the \ptr\ distribution
from simulation is scaled to have an integral equal to the number of events 
observed in data.  For each trial, the number of events in each bin is fluctuated 
according to a Poisson distribution with a mean set by the original bin content.
The unfolding procedure is used on the fluctuated distribution, and the \ptw\ 
distribution from the same set of simulated events is subtracted from the result.
The resulting ensemble of offsets is used to fill a covariance matrix describing 
the impact of statistical fluctuations on the result, including correlations between
the bins introduced by the unfolding procedure.

Systematic uncertainties receive contributions from
the quality of the response parametrization approximation, $i.e.$ from the difference between $M_{\rm{MC}}$ and $M_{\rm{param}}$; from
the statistical precision of the data driven corrections defining $M_{\rm{param}}^{\rm{corr}}$; and from the
unfolding procedure itself. Their estimation is described in Section~\ref{sec:Systematics}. 

\subsection{\label{sec:Efficiency}Efficiency Correction}

The \Wln\ candidate event reconstruction efficiency is
subsequently unfolded by dividing the number of events in each bin of
\ptw\ by the detection efficiency correction factor for that bin.
The correction factor accounts for trigger and detection efficiencies, as well as 
the migration of events in and out of the acceptance due to charged lepton and \met\ 
resolution effects.  It is defined
as the ratio of the number of reconstructed events passing all selection in each
bin to the number of events produced within the fiducial volume in that same bin.
Note that any migration between bins has already been accounted for by the hadronic
recoil response unfolding.
The efficiency correction is based on the ratio calculated from simulated $W$ events, 
and corrected for observed differences between simulated and real data
in the trigger and reconstruction efficiencies as well as in the lepton 
momentum and resolution (see Section~\ref{sec:MCSamples}).  The corrections for 
discrepancies between data and  simulation are applied as a function of the 
reconstructed lepton kinematics in each bin of \ptw.
The fiducial volume in the denominator is defined by the truth-level 
kinematic requirements $p_T^\ell>20$~GeV, $|\eta_\ell|<2.4$,
$p_T^\nu>25$~GeV, and $m_T>40$~GeV. For the default, propagator-level \ptw\ 
measurement, the lepton kinematics and transverse mass are defined at the QED 
Born level, \emph{i.e.},  before any final state QED radiation.  For the ``dressed'' 
lepton version of the measurement, the charged lepton momentum is the sum of 
its momentum after all QED FSR and the momenta of all photons radiated within 
a cone of $\Delta R = 0.2$ around the lepton.  
The cone size is chosen to match the cone size used for the lepton removal
in the definition of \hr.  The ``bare'' lepton version uses 
only the charged lepton momentum after all QED FSR.

In the electron channel, the efficiency rises from~ $\sim 60$\% at
low \ptw\ to $\sim 80$\% at $\ptw\sim 100$~GeV, and falls towards
$\sim 70$\% at the upper end of the spectrum. In the muon channel,
the efficiency rises from $\sim 80$\% to $\sim 90$\%, then falls
to $\sim 80$\% in the same \ptw\ ranges. 

The efficiency correction carries systematic uncertainties induced by
the imperfect modeling of the lepton trigger and reconstruction
efficiencies, by the acceptance of the \met\ cut, and by the finite
statistics and physics assumptions of the signal simulation
sample. Their estimation is described in Section~\ref{sec:Systematics}.

\section{\label{sec:Systematics}Systematic Uncertainties}

Systematic uncertainties arise from the background subtraction
procedure, from the recoil response model and unfolding procedure, and
from lepton reconstruction and calibration uncertainties. Theoretical
uncertainties also enter, to a lesser extent.  Different
strategies are used for the various uncertainties according to the
nature of the uncertainty and whether it is introduced before, during,
or after the hadronic recoil unfolding.  Accordingly, the uncertainties
are evaluated by using an ensemble of inputs with the nominal response 
matrix, an ensemble of response matrices with the nominal input, or by 
simple error propagation, respectively.  The uncertainties on this measurement
are represented as covariance matrices, so that correlations between the bins
can be included.  

\subsection{Background Subtraction Uncertainties}

The systematic uncertainties associated to the background subtraction
are estimated by generating an ensemble of pseudo-experiments in which
the background estimates have been fluctuated within their uncertainties.
The full analysis chain is repeated for
each pseudo-experiment and the spread of the unfolded results defines
the associated uncertainty. Electroweak, top, and QCD multijet
contributions are treated separately, 
except that the luminosity uncertainty is treated as correlated
between the electroweak and top backgrounds. 		
Background subtraction is performed before the unfolding, and the 
unfolding redistributes the background among the \ptw\ bins, so the 
covariance matrices representing the uncertainties on the backgrounds 
have nonzero off-diagonal elements.

The electroweak and top backgrounds contribute 0.6\% (0.4\%) to the
measurement uncertainty at low \ptw\ in the electron (muon) channel,
and up to 4\% at high \ptw\ in both channels.
The multijet background in the electron channel contributes $\sim 0.5$\%
uncertainty for $\ptw<50$~GeV, which gradually rises to 4\% at $\ptw\sim
200$~GeV, eventually contributing 15\% in the highest \ptw\ bin. In the muon
channel, the multijet background induced uncertainty has a maximum of 2\%
at $\ptw\sim 30$~GeV, which corresponds to the peak of the 
background rate, and contributes $\sim 0.6$\% on average in the rest
of the spectrum.

\subsection{Hadronic Recoil Unfolding Uncertainties}

Systematic uncertainties associated to the response matrix are
classified in two categories. In the first category, the impact of a
given source of uncertainty is estimated by comparing the unfolded
distribution obtained with the nominal response matrix, to the result
obtained with a response matrix reflecting the variation of this
source. The statistical component of the difference is assessed by
varying a given input of the response matrix construction to generate 
a set of related variations of the response matrices.
Repeating the analysis with these leads to a set of
varied unfolded results, and the induced bias is averaged in
each bin of the \ptw\ distribution. The associated systematic
uncertainty is defined from the spread of the distribution of the
results, and is taken as a constant percentage across all \ptw\ bins,
represented as a diagonal covariance matrix. 

By comparing results obtained from the initial Monte Carlo response
matrix $M_{\rm{MC}}$ with results obtained from the parametrized response 
matrix $M_{\rm{param}}$,  
the response parametrization is found to induce an uncertainty of
2.4\% in the electron channel and 2.0\% in the muon
channel. The input generator bias is estimated by reweighting the true
\ptw\ distribution given by the \pythia\ sample to the 
\resbos\ prediction, generating the corresponding response matrix and
comparing the result to the nominal result, leading to a systematic
uncertainty of 1.2\% in the electron channel, and 0.9\% in the muon channel. 
Note that the starting assumption for the Bayesian unfolding is	
simultaneously modified in the same way, so that this uncertainty
includes both the effect of modifying the distribution underlying 
the response matrix and the assumption of a prior for the unfolding.
In addition, it was verified that reweighting the input
\ptw\ assumption according to the actual measurement result and
repeating the procedure does not affect the result beyond the
uncertainties quoted above.
Lepton momentum scale uncertainties also enter through the
$Z$-based recoil response corrections, because \ptll\ is used in place of
the true \ptz, but this amounts to less than 0.2\% in
both channels. As described above, these numbers are taken constant across
the \ptw\ spectrum.

The second category deals with the uncertainties associated to the
data-driven corrections to the response parametrization. In this case, we generate an ensemble of
random correction parameters by sampling from the distribution defined by the
statistical uncertainties on the central value of the parameters returned by the fit.
For each parameter set the corresponding response
matrix is generated. The treatment is then the same as for the
background uncertainties: the analysis chain is repeated for
each configuration, and the spread of the unfolded bin contents
defines the associated uncertainty in each bin.  

In this category, the data driven correction to the recoil bias and
resolution induces an uncertainty of $\sim 1.6$\% for $\ptw<8$~GeV,
has a local maximum of $\sim 2.6$\% at $\ptw=30$~GeV, and contributes
less than 1\% in the remaining part of the spectrum. The uncertainty
related to the \sumet\ rescaling is 0.2\% at low \ptw, rising to 1\%
at the high end of the spectrum. These numbers are valid for both
channels, as the data driven corrections are determined from combined
$Z\rightarrow ee$ and $Z\rightarrow\mu\mu$ samples, as described in
Section~\ref{sec:Unfolding}.

Finally, the bias from the unfolding itself is found by folding
the \ptw\ distribution of simulated \Wln\ events passing the reconstruction-level selection
using $M_{\rm{MC}}$ and then unfolding it
using the same response matrix.  The original \ptw\ distribution is 
subtracted from the unfolded one, and the size of the bias relative to the  
original distribution is taken as the systematic uncertainty from the 
unfolding procedure.  The folded distribution is used for \ptr\ instead
of the found \ptr\ distribution to avoid double-counting the statistical
uncertainty.  The resulting uncertainty is less than 0.5\% in all bins, 
except for the highest-\ptw\ bin in the electron channel, where it is 1\%.

\subsection{Efficiency Correction Uncertainties}

In the electron channel, the main contributions to the acceptance
correction uncertainty are the reconstruction and identification
efficiency uncertainty, and the electron energy scale and resolution
uncertainties. The identification efficiency contributes 3\% to the
measurement uncertainty across the \ptw\ spectrum. The scale and
resolution uncertainties contribute 0.5\% at low \ptw, rising to 10\%
at $\ptw\sim 100$~GeV, and decreasing to 6\% at the high end of the
spectrum.

In the muon channel, the trigger efficiency uncertainty contributes
1\% across the spectrum. The reconstruction efficiency contributes
0.7\% at low \ptw, linearly rising to 2\% at $\ptw\sim 300$~GeV.
The scale and resolution uncertainties contribute 0.5\% at low \ptw,
rising to 2\% at $\ptw\sim 120$~GeV, and decreasing to 1\% towards
$\ptw\sim 300$~GeV.

The uncertainty associated to the recoil component of \met\ 
(the first term of Eq.~\ref{eq:met}, minus any clusters associated with an electron)
is estimated as above, by generating random ensembles of resolution
correction parameters within the precision of the $Z$-based
calibration.  For each parameter set in the ensemble, the \met\ distribution 
is re-generated and the corresponding efficiency correction is re-calculated. 
The width of the resulting distribution of efficiency corrections
is taken as the uncertainty. This source contributes
less than 0.3\% across the \ptw\ spectrum in both channels.

In both channels, the Monte Carlo statistical precision is 0.5\% at
low \ptw\ and rises to 4\% towards $\ptw \sim 300$~GeV.
The generator dependence of the efficiency is estimated by comparing the
central values found for \pythia\ and \mcatnlo, and
found to be smaller than $0.2$\%, apart from the
last bin where it reaches 1\%.  Finally, following Ref.~\cite{ZpTPaper}, the PDF
induced uncertainty on the efficiency correction is at the level
of 0.1\% and neglected in this analysis.

\section{\label{sec:Results}Results}

\subsection{Electron and Muon Channel Results}

The efficiency-corrected distributions resulting from the two unfolding steps
are normalized to unity, and the bin contents are divided by the bin width. 
In the normalization step, uncertainties that are completely correlated across all of
the bins, such as the uncertainty on the integrated luminosity, cancel.
The resulting normalized differential fiducial cross section, \dsdptwNorm\
is given in Table~\ref{tab:results1} for both the electron and muon channels, 
together with the statistical and systematic uncertainties. The differential
cross section is calculated with respect to three definitions of \ptw\ and 
the fiducial volume, corresponding to different definitions of the true lepton kinematics:
the first uses the Born level kinematics, 
the second uses the ``dressed'' lepton kinematics calculated
from the sum of the post-FSR lepton momentum and the momenta of all
photons radiated within a cone of $\Delta R = 0.2$,  
and the third (``bare'') uses the lepton kinematics after all QED radiation.

\begin{table*}[h]
\caption{\label{tab:results1}%
    The normalized, differential cross section \dsdptwNorm,
    measured in \Wen\ and \Wmn\ events, for different definitions of \ptw.
    The Born-level definition (``propag.''), 
    the analysis baseline, ignores the leptons and takes the $W$ momentum from the
    propagator.  The ``dressed'' and ``bare'' definitions of \ptw\ are
    calculated using the momenta of the leptons from the $W$ decay.  
    In the ``dressed'' case, the charged lepton momentum
    includes the momenta of photons radiated within a cone of 
    $\Delta R = 0.2$ centered around the lepton.  In the ``bare'' case, the 
    charged lepton momentum after all QED radiation is used. The factor 
    $p$ is the power of ten to be multiplied by each of the three cross section
    numbers for each channel.  It has been factorized out for legibility.
}
\begin{ruledtabular}
\begin{tabular}{D{-}{-}{-1}dddcddcdddcdd}
\multicolumn{1}{c}{\ptw\ Bin}&\multicolumn{13}{l}{\dsdptwNorm (GeV$^{-1})$}\\
\multicolumn{1}{c}{[GeV]}&\multicolumn{4}{l}{\Wen}&\multicolumn{2}{c}{uncert. (\%)}& &\multicolumn{4}{l}{\Wmn}&\multicolumn{2}{c}{uncert. (\%)}\\
\cline{2-7}\cline{9-14}
\multicolumn{1}{c}{
}&\multicolumn{1}{c}{propag.}&\multicolumn{1}{c}{dressed}&\multicolumn{1}{c}{bare}&\multicolumn{1}{c}{$p$}&\multicolumn{1}{c}{stat.}&\multicolumn{1}{c}{syst.}& &\multicolumn{1}{c}{propag.}&\multicolumn{1}{c}{dressed}&\multicolumn{1}{c}{bare}&\multicolumn{1}{c}{$p$}&\multicolumn{1}{c}{stat.}&\multicolumn{1}{c}{syst.}\\
\hline
0\,-\,8  & 5.60 & 5.55 & 5.42 & $\,10^{-2}$ & 0.4 & 2.8 & & 5.44 & 5.39 & 5.35 & $\,10^{-2}$ & 0.4 & 2.6 \\
8\,-\,23  & 2.50 & 2.52 & 2.56 & $\,10^{-2}$ & 0.4 & 2.9 & & 2.52 & 2.54 & 2.55 & $\,10^{-2}$ & 0.3 & 2.6 \\
23\,-\,38  & 6.66 & 6.76 & 6.96 & $\,10^{-3}$ & 0.9 & 4.7 & & 6.96 & 7.06 & 7.11 & $\,10^{-3}$ & 0.8 & 4.7 \\
38\,-\,55  & 2.46 & 2.46 & 2.46 & $\,10^{-3}$ & 1.3 & 4.8 & & 2.55 & 2.55 & 2.55 & $\,10^{-3}$ & 1.3 & 4.0 \\
55\,-\,75  & 9.39 & 9.35 & 9.19 & $\,10^{-4}$ & 2.0 & 7.4 & & 1.04 & 1.04 & 1.03 & $\,10^{-3}$ & 2.0 & 3.9 \\
75\,-\,95  & 3.75 & 3.73 & 3.64 & $\,10^{-4}$ & 3.4 & 9.5 & & 4.40 & 4.37 & 4.34 & $\,10^{-4}$ & 3.3 & 4.1 \\
95\,-\,120  & 1.82 & 1.80 & 1.75 & $\,10^{-4}$ & 4.1 & 10.8 & & 1.92 & 1.90 & 1.88 & $\,10^{-4}$ & 4.4 & 4.9 \\
120\,-\,145  & 9.56 & 9.49 & 9.19 & $\,10^{-5}$ & 6.0 & 10.1 & & 7.35 & 7.29 & 7.21 & $\,10^{-5}$ & 7.5 & 6.4 \\
145\,-\,175  & 3.57 & 3.54 & 3.43 & $\,10^{-5}$ & 7.9 & 10.4 & & 3.99 & 3.96 & 3.91 & $\,10^{-5}$ & 11.0 & 5.8 \\
175\,-\,210  & 1.59 & 1.58 & 1.52 & $\,10^{-5}$ & 10.0 & 8.9 & & 1.88 & 1.86 & 1.84 & $\,10^{-5}$ & 14.7 & 7.4 \\
210\,-\,300  & 4.71 & 4.67 & 4.49 & $\,10^{-6}$ & 12.2 & 15.5 & & 4.68 & 4.66 & 4.55 & $\,10^{-6}$ & 17.9 & 13.1 \\
\end{tabular}
\end{ruledtabular}
\end{table*}

\begin{figure}
    \centering
    \includegraphics[width=0.49\textwidth]{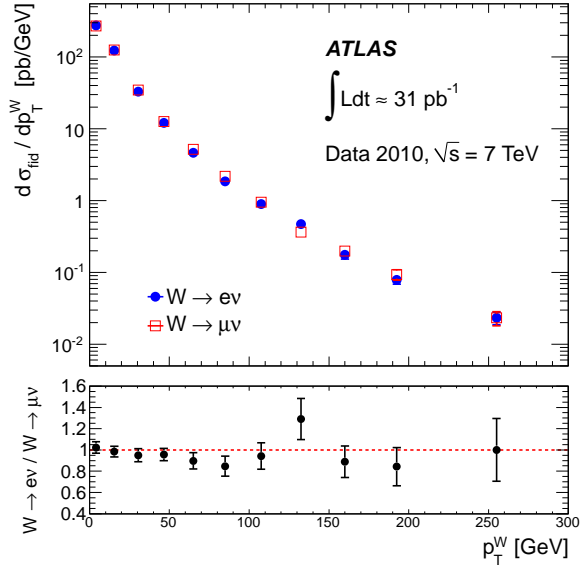}
    \caption{\label{fig:BothChannelsFid}
	Electron and muon fiducial differential cross sections as a
        function of \ptw. The error bars include all statistical and
        systematic uncertainties except the 3.4\% uncertainty on the integrated
	luminosity, which is common to the two measurements and
        cancels in the ratio.} 
\end{figure}

Instead of normalizing the efficiency-corrected distributions to unit integral,
they can also be divided by the integrated luminosity of the corresponding 
data to yield the differential fiducial cross section \dsdptw.  
The resulting differential fiducial cross sections, with the fiducial volume
defined by the Born-level kinematics, are shown in Fig.~\ref{fig:BothChannelsFid}. Error
bars include both statistical and systematic uncertainties, but not the uncertainty on 
the integrated luminosity, which is common to both measurements.

\subsection{\label{sec:combination}Combination Procedure}

After correcting the electron and muon \ptw\ distributions to the common fiducial
volume using the efficiency corrections described in 
Section~\ref{sec:HadRecUnfold}, 
we combine the resulting differential fiducial cross sections \dsdptw\ 
by $\chi^2$ minimization.  The combination is based on the distributions with \ptw\ 
defined by the $W$ propagator momentum because QED final state radiation causes 
differences between the electron and muon momenta that makes a consistent combination 
based on other definitions unfeasible.
To build the $\chi^2$, the uncertainties on the two measurements are sorted according 
to whether they are correlated between the two channels or not, and a joint covariance 
matrix describing the uncertainty on both measurements is constructed.  Using this 
covariance matrix, we define a $\chi^2$ between the two measurements and a common 
underlying distribution.  This $\chi^2$ is minimized to find the combined measurement,   
which is the best estimate of the common underlying distribution.

Specifically, the $\chi^2$ to be minimized is defined as:
\begin{equation}
\chi^2 = \bf (X-\bar{X})^T C^{-1} (X-\bar{X}) \,\,,
\label{eq:chisquared}
\end{equation}
where ${\bf X}$ is the vector of $2N$ elements containing the two $N$-bin distributions
to be combined, concatenated: ${\bf X} = \{X^e_1,...,X^e_n;X^\mu_1,...,X^\mu_n\}$.  
The vector ${\bf \bar{X}} = \{\bar{X}_1,...,\bar{X}_n;\bar{X}_1,...,\bar{X}_n\}$ 
contains two copies of the combined measurement $\{\bar{X}_i\}$.  The joint 
covariance matrix 
${\bf C}$ is described in the next paragraph. The $\chi^2$ minimization is
performed analytically, following the prescription in Ref.~\cite{PDG},
yielding the $\{\bar{X}_i\}$.

The joint covariance matrix ${\bf C}$ has $2N \times 2N$ elements and is 
constructed from four sub-matrices:
\begin{equation}
{\bf C} = \left( \begin{array}{cc}
C^e & C^{e\mu} \\
C^{e\mu} & C^\mu \\
\end{array} \right)  \,\,.
\label{eq:covariance}
\end{equation}
The $N \times N$ covariance matrices $C^e$ and $C^\mu$ are the covariance
matrices for the electron and muon measurements, respectively, and contain all
sources of uncertainty on the measurements.  The off-diagonal blocks $C^{e\mu}$ 
are identical and reflect the sources of uncertainty that are correlated between the channels.

The $2N \times 2N$ covariance matrix is constructed from the two 
$N \times N$ matrices for each source of uncertainty individually, and the resulting set of 
$2N \times 2N$ matrices is summed.  For sources of uncertainty uncorrelated 
between the channels, the $2N \times 2N$ covariance matrix is constructed by copying 
the $N \times N$ matrices to the corresponding diagonal blocks $C^e$ and $C^\mu$.
For uncertainties that are correlated between the channels, the diagonal blocks
are still filled by copying the covariance matrices from the individual channels.
The off-diagonal blocks are filled using the assumption that the channels are 
100\% correlated, so that the correlations between bins are identical for both
channels.  That determines the correlation matrix, which sets the magnitudes of the 
covariance matrix entries relative to the magnitude of the diagonal entries.
The diagonal entries, which are the squares on the uncertainties on each bin, are 
taken as the geometrical average of the values for the two channels.

The statistical uncertainties on the unfolded measurements are uncorrelated because
the \Wen\ and \Wmn\ candidate data samples are statistically independent.  The 
systematic uncertainties induced by the subtraction of the estimated background are
uncorrelated between the channels, except for the uncertainties on the luminosity and
predicted cross sections used to normalize the electroweak and top quark backgrounds.
Because the same hadronic recoil response matrix is used for both channels, the 
uncertainties associated with it are fully correlated between the channels, except
for the small contribution from the lepton momentum resolution.  The
efficiency corrections for each channel are independent, so the associated uncertainties
are uncorrelated between the channels.

\subsection{Combined Results and Comparison with Predictions}

\begin{figure}
    \centering
    \includegraphics[width=0.49\textwidth]{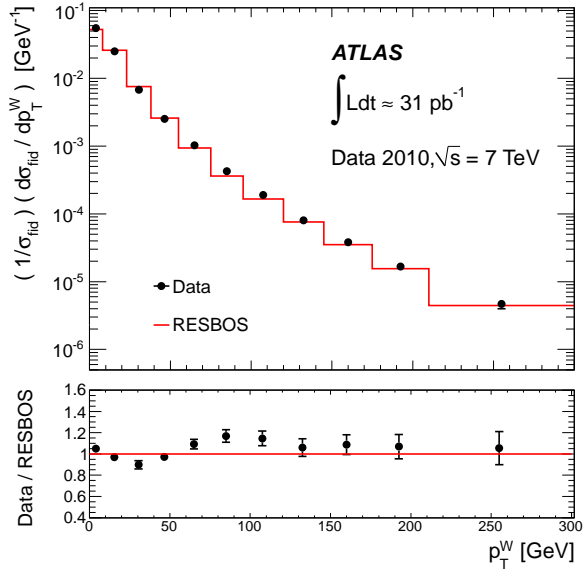}
    \caption{\label{fig:CombinedResultNorm}
      Normalized differential cross section obtained from the combined
      electron and muon measurements, compared to the \resbos\
      prediction.}
\end{figure}

The $\chi^2$ minimization yields a $\chi^2$/d.o.f.~of 13.0/13, demonstrating
good agreement between the electron and muon results.
The combined differential cross section, normalized to unity, is shown compared
to the prediction from \resbos\ in Fig.~\ref{fig:CombinedResultNorm}.
The \resbos\ prediction, which combines resummed and fixed order pQCD calculations, is based on
the CTEQ6.6 PDF set~\cite{cteq66} and a renormalization and factorization scale of $m_W$. 
\resbos\ performs the fixed-order calculation at NLO ($\mathcal{O}(\alpha_s)$), 
and corrects the prediction to NNLO ($\mathcal{O}(\alpha_s^2)$)
using a $k$-factor calculated as a function of the boson mass, rapidity, 
and $p_T$~\cite{resbos1,resbos4,resbos20}.  
Table~\ref{tab:combResult} gives the same information numerically, including the 
separate contribution of different classes of uncertainty.
\begin{table*}[h]
\caption{\label{tab:combResult}%
Measured \ptw\ using combined electron and muon data, with all uncertainties
shown by source. }
\begin{ruledtabular}
\begin{tabular}{D{-}{-}{-1}|D{x}{\cdot}{-1}|dddd|d}
\multicolumn{1}{c}{\ptw\ Bin}&
\multicolumn{1}{c}{\dsdptwNorm}&
\multicolumn{1}{c}{ResponseMatrix}&
\multicolumn{1}{c}{Backgrounds}&
\multicolumn{1}{c}{Efficiency}&
\multicolumn{1}{c}{Statistical}&
\multicolumn{1}{c}{Total}\\
\multicolumn{1}{c}{[GeV]}&
\multicolumn{1}{c}{(GeV$^{-1}$)}&
\multicolumn{1}{c}{uncert. (\%)}&
\multicolumn{1}{c}{uncert. (\%)}&
\multicolumn{1}{c}{uncert. (\%)}&
\multicolumn{1}{c}{uncert. (\%)}&
\multicolumn{1}{c}{uncert. (\%)}\\
\hline
0\,-\,8  & 5.510\,x\,10^{-2} & 1.91 & 0.26 & 0.76 & 0.22 & 2.48 \\
8\,-\,23  & 2.512\,x\,10^{-2} & 1.69 & 0.28 & 0.87 & 0.24 & 2.42 \\
23\,-\,38  & 6.766\,x\,10^{-3} & 3.20 & 0.57 & 1.28 & 0.57 & 4.31 \\
38\,-\,55  & 2.523\,x\,10^{-3} & 2.34 & 0.65 & 1.44 & 0.84 & 3.78 \\
55\,-\,75  & 1.025\,x\,10^{-3} & 1.78 & 0.74 & 1.74 & 1.19 & 4.09 \\
75\,-\,95  & 4.263\,x\,10^{-4} & 1.61 & 1.15 & 2.13 & 1.91 & 4.94 \\
95\,-\,120  & 1.896\,x\,10^{-4} & 1.98 & 1.94 & 2.67 & 2.68 & 5.99 \\
120\,-\,145  & 7.985\,x\,10^{-5} & 2.84 & 3.30 & 3.16 & 4.78 & 7.91 \\
145\,-\,175  & 3.710\,x\,10^{-5} & 1.98 & 2.66 & 3.66 & 5.72 & 9.31 \\
175\,-\,210  & 1.692\,x\,10^{-5} & 2.00 & 3.72 & 3.84 & 7.75 & 10.56 \\
210\,-\,300  & 4.803\,x\,10^{-6} & 2.69 & 7.81 & 4.26 & 9.28 & 14.40 \\
\end{tabular}
\end{ruledtabular}
\end{table*}

\begin{figure*}[tbp]
  \begin{overpic}[width=0.49\textwidth]{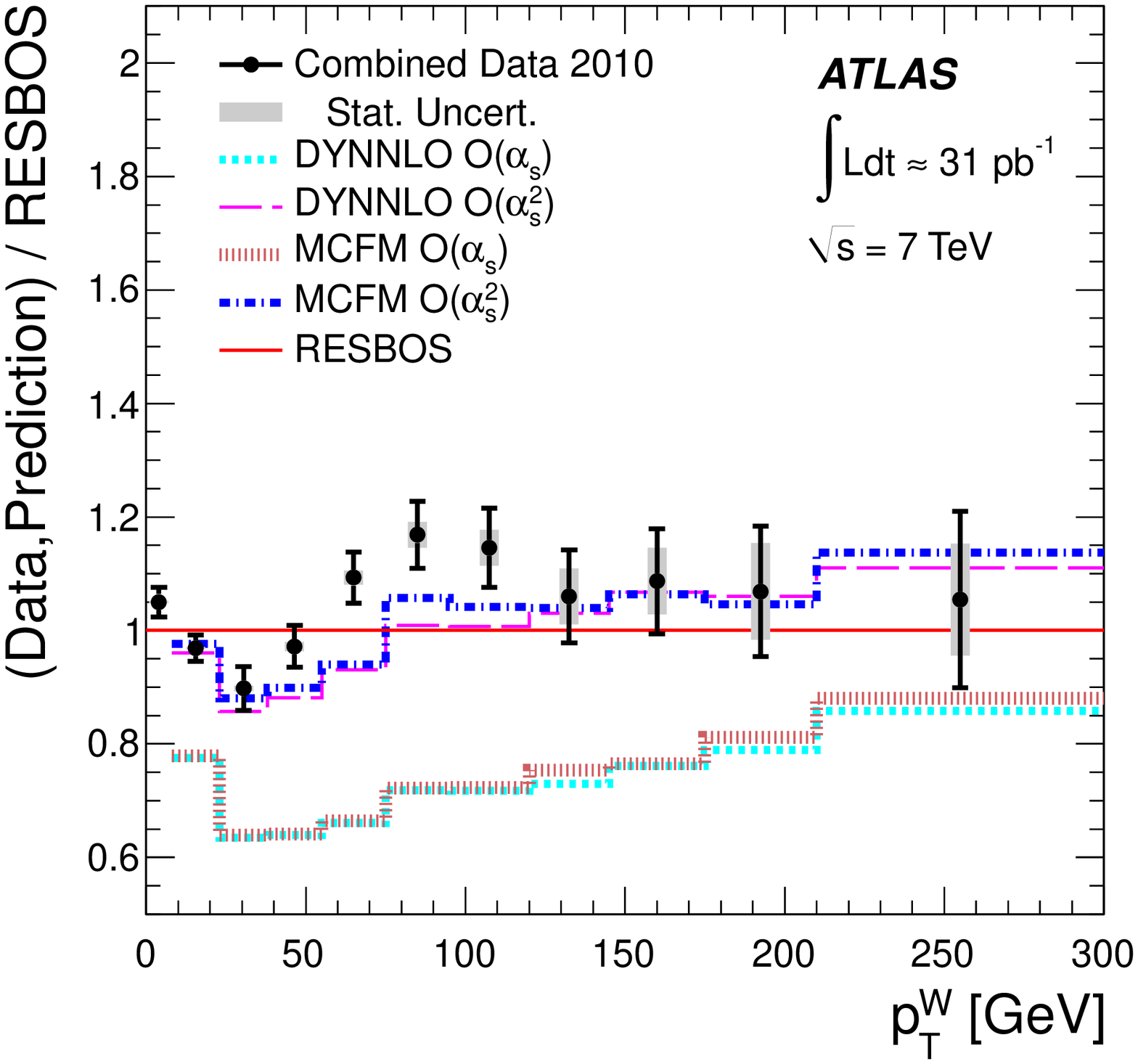}
    \put(80,20){(a)}
  \end{overpic}
  \begin{overpic}[width=0.49\textwidth]{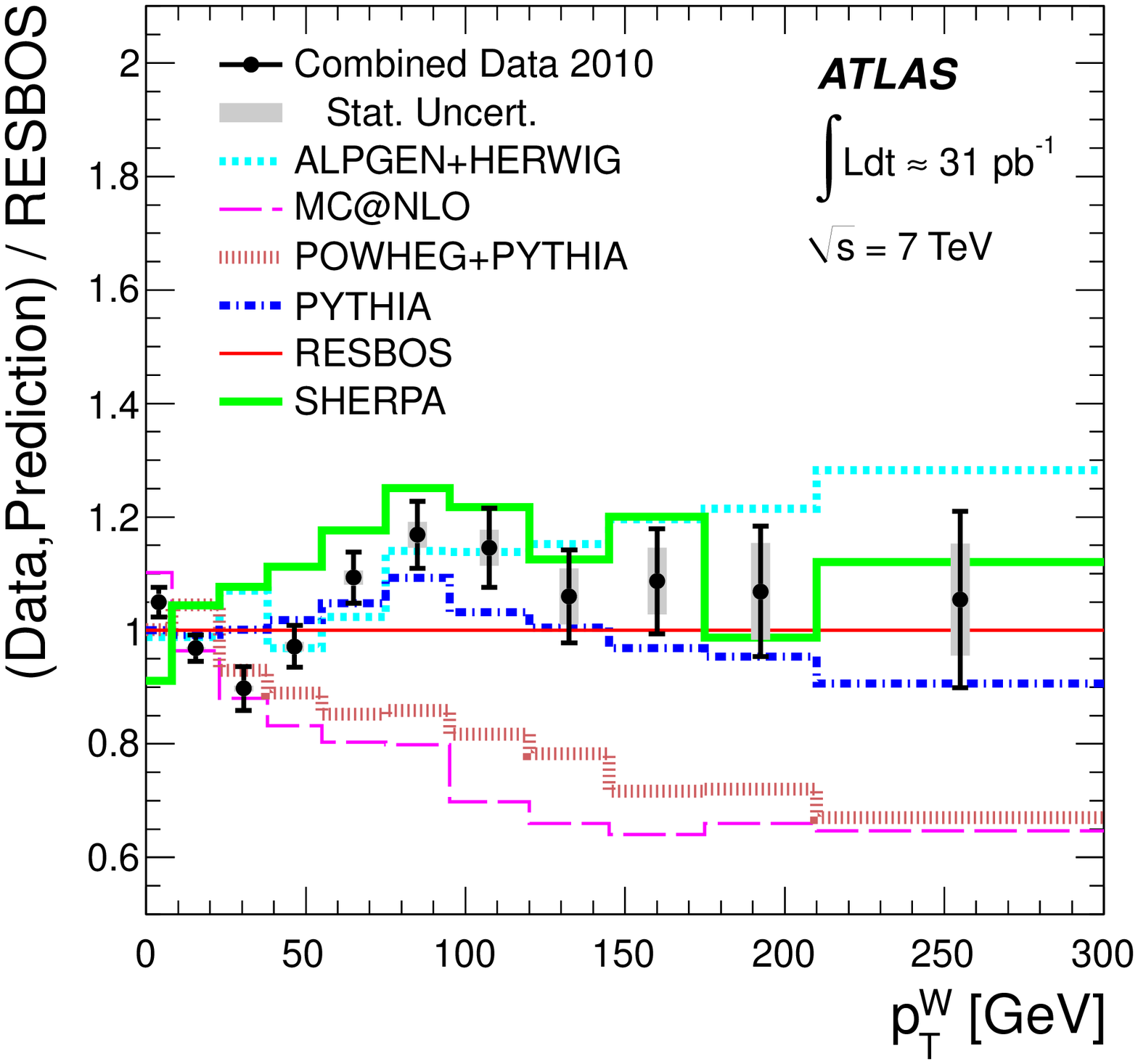}
    \put(20,20){(b)}
  \end{overpic}
\caption{\label{fig:TheoryComp} 
    Ratio of the combined measurement and various predictions to the \resbos\ prediction for \dsdptwNorm, 
    using (a) the $\mathcal{O}(\alpha_s)$ and $\mathcal{O}(\alpha_s^2)$ predictions from \dynnlo\ and 
    \mcfm, and using (b) the predictions from \alpgen+\herwig, \mcatnlo, \powheg+\pythia, \pythia, and \sherpa. 
    The statistical uncertainties on the generator distributions are negligible compared to the 
    uncertainty on the measurement and are not shown.} 
\end{figure*}

In Figure~\ref{fig:TheoryComp}, the combined result \dsdptwNorm\ is 
compared to a selection of predictions from both pQCD and event generators.
The \dynnlo\ predictions are from version 1.1 of the program~\cite{dynnlo,dynnlo2}.
The prediction from the \mcfm\ program is produced as a calculation of 
\dsdptw\ for $W$~+~1 parton events and uses \mcfm\ version 5.8~\cite{mcfm}.
The leading order calculation for $W$~+~1 parton production is $\mathcal{O}(\alpha_s)$  
and the NLO calculation is $\mathcal{O}(\alpha_s^2)$, so the predictions are 
comparable to other $\mathcal{O}(\alpha_s)$ and $\mathcal{O}(\alpha_s^2)$ 
predictions of \ptw\ for $\ptw > 5$~GeV, the minimum jet $p_T$ threshold
in the calculation.
Both of the pQCD calculations are normalized by dividing the prediction in each bin by the
inclusive cross section prediction calculated in the same configuration as the differential 
cross section, and both have the renormalization and factorization scales set to $m_W$.
The $\mathcal{O}(\alpha_s)$ predictions use the MSTW2008 NLO PDF sets, and the 
$\mathcal{O}(\alpha_s^2)$ predictions use the NNLO MSTW2008 PDF set~\cite{mstw2008}.   
The uncertainty on the pQCD predictions comes mostly from the renormalization and
factorization scale dependence, and studies indicate that it is 
comparable in magnitude to the 10\% and 8\% observed for \ptz\ predictions at 
$\mathcal{O}(\alpha_s)$ and $\mathcal{O}(\alpha_s^2)$ in Ref.~\cite{ZpTPaper}.

The \dynnlo\ and \mcfm\ predictions do not include resummation effects and are
not expected to predict the data well at low \ptw\ because of the diverging 
prediction for vanishing \ptw.  Therefore, the lowest bin ($\ptw < 8$~GeV) 
is omitted from Fig.~\ref{fig:TheoryComp}.
The two programs predict similar distributions at the same order of $\alpha_s$.
The $\mathcal{O}(\alpha_s)$ prediction from both calculations 
for the fraction of the distribution above $\ptw\sim 23$~GeV is about 30\% too 
low on average, similar to the NLO event generators.  The $\mathcal{O}(\alpha_s)$ 
prediction from \fewz~\cite{fewz1,fewz2} is not shown in Fig.~\ref{fig:TheoryComp}
but is in agreement with those from \dynnlo\ and \mcfm.
The discrepancy between the predictions and the measurement appears when normalizing to 
the inclusive cross section and would be compensated by a large but unphysical contribution 
in the first bin.  The ratio moves closer to unity in the high \ptw\ range. 
The $\mathcal{O}(\alpha_s^2)$ predictions agree better with the data 
than those at $\mathcal{O}(\alpha_s)$.  They are within 15\% of the data for all \ptw.

The predictions of the event generators \pythia, \powheg, \alpgen, \sherpa, and \mcatnlo\
are based on the simulated samples described in Section~\ref{sec:MCSamples}. 
Since \powheg\ and \alpgen\ can be interfaced with more than one
parton shower implementation, the notations \powheg+\pythia\ and
\alpgen+\herwig\ are used to make the choice explicit. The \pythia,
\resbos, \sherpa, and \alpgen+\herwig\ predictions describe the measurement
within 20\% over the entire range. For $\ptw<38$~GeV, the data
indicate a softer spectrum than these predictions. For
$38<\ptw<120$~GeV, the data distribution exceeds the
\resbos\ prediction and undershoots the \sherpa\ prediction,  but
agrees with the \alpgen+\herwig\ and, to a lesser extent,
pure \pythia\ predictions. For $\ptw>120$~GeV, \pythia\ and \resbos\ agree
in predicting a softer spectrum than \alpgen+\herwig\ and \sherpa, but the
data provide no significant discrimination among these predictions. 

\powheg+\pythia\ and \mcatnlo, the NLO event generators interfaced with parton 
shower algorithms, provide a reasonable
description of the data for $\ptw<38$~GeV, but both underestimate
the data starting at $\ptw \approx 38$~GeV, with a deficit gradually
increasing to nearly 40\% at high \ptw. 

\begin{figure}
\begin{center} 
\includegraphics[width=0.49\textwidth]{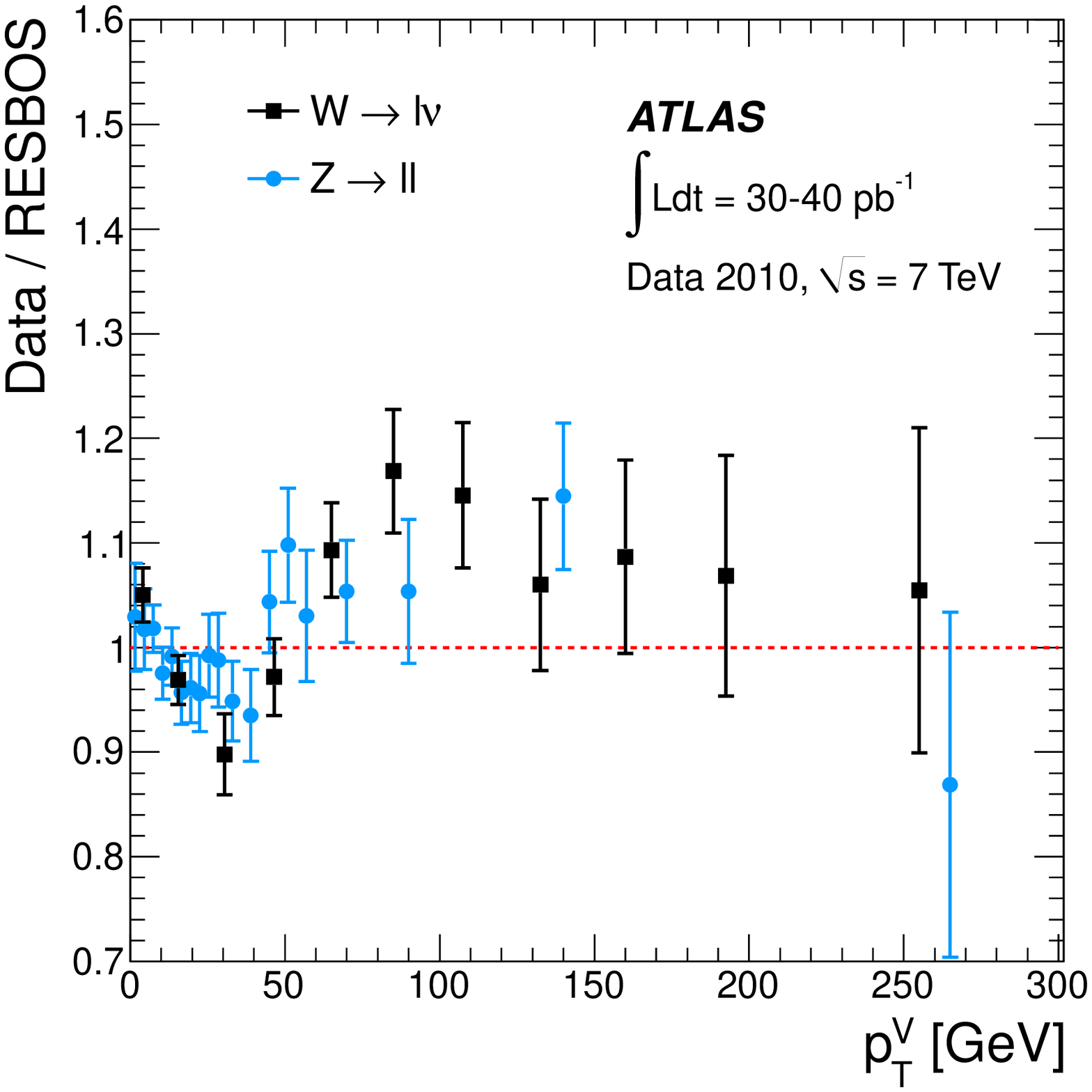}
\end{center} 
\caption{The ratio of \dsdptwNorm\ as measured in the combined  
    electron and muon data to the \resbos\ prediction, overlaid with the  
    ratio of \dsdptzNorm\ measured as described in Ref.~\cite{ZpTPaper} 
    to the \resbos\ prediction.} 
    \label{fig:CompareZpTWpT} 
\end{figure}

Finally, we compare the combined result to the measurement of \dsdptzNorm\
described in Ref.~\cite{ZpTPaper}.  
The $W$ and $Z$ have different masses and couple differently to quarks, so
the results cannot be directly compared, but the ratios of the measured to 
predicted distributions for  
a common model can be used to qualitatively assess the agreement between the two 
measurements.  The ratios of the $W$ and $Z$ distributions in data to 
their respective \resbos\ predictions are overlaid in Fig.~\ref{fig:CompareZpTWpT}.   
In spite of the different techniques and uncertainties characterizing
both measurements, the ratios display similar trends as a function of \ptv, the 
true boson $p_T$.

\section{\label{sec:Conclusions}Conclusions} 

The $W$ transverse momentum differential cross section has been measured
for \mbox{$\ptw < 300$ GeV} in \Wln\ events reconstructed in the electron and 
muon channels using the ATLAS detector.  The \Wln\ candidate events are selected from 
$pp$ collision data produced at $\sqrt{s}=7$~TeV, corresponding 
to approximately 31~\ipb\ from the 2010 run of the LHC.  

The measurement is compared to a selection of predictions. The \alpgen+\herwig, 
\pythia, \resbos, and \sherpa\ predictions match the data within 20\% over the entire
\ptw\ range. \mcatnlo\ provides the closest description of the data for
$\ptw<38$~GeV, but \mcatnlo\ and \powheg+\pythia\ both underestimate the data at higher \ptw. 
Fixed-order pQCD predictions from the \dynnlo\ and \mcfm\ programs
agree very well with each other.  They predict fewer events at high 
\ptw\ at $\mathcal{O}(\alpha_s)$ but the agreement with the measured distribution 
is significantly improved by the $\mathcal{O}(\alpha_s^2)$ calculations.

A comparison of the $W$ and $Z$ data relative to the prediction from a given
theoretical framework displays similar features across the measured
transverse momentum range, supporting the expected universality of strong
interaction effects in $W$ and $Z$ production.

Although the measurement is limited by systematic uncertainties over
most of the spectrum, the dominant uncertainty sources can be 
constrained with more integrated luminosity.  With the  integrated
luminosity available from the 2011 run now in  progress, future
measurements should be able to measure \dsdptw\ to at least double the
current range in \ptw. With improved statistical and systematic
uncertainties, it should also be possible to measure the ratios of the
$W$ to $Z$ and $W^+$ to $W^-$  differential cross sections as
functions of the boson $p_T$, which will further test the
predictions of QCD.

\section{Acknowledgements}

We thank CERN for the very successful operation of the LHC, as well as the
support staff from our institutions without whom ATLAS could not be
operated efficiently.

We acknowledge the support of ANPCyT, Argentina; YerPhI, Armenia; ARC,
Australia; BMWF, Austria; ANAS, Azerbaijan; SSTC, Belarus; CNPq and FAPESP,
Brazil; NSERC, NRC and CFI, Canada; CERN; CONICYT, Chile; CAS, MOST and
NSFC, China; COLCIENCIAS, Colombia; MSMT CR, MPO CR and VSC CR, Czech
Republic; DNRF, DNSRC and Lundbeck Foundation, Denmark; ARTEMIS, European
Union; IN2P3-CNRS, CEA-DSM/IRFU, France; GNAS, Georgia; BMBF, DFG, HGF, MPG
and AvH Foundation, Germany; GSRT, Greece; ISF, MINERVA, GIF, DIP and
Benoziyo Center, Israel; INFN, Italy; MEXT and JSPS, Japan; CNRST, Morocco;
FOM and NWO, Netherlands; RCN, Norway; MNiSW, Poland; GRICES and FCT,
Portugal; MERYS (MECTS), Romania; MES of Russia and ROSATOM, Russian
Federation; JINR; MSTD, Serbia; MSSR, Slovakia; ARRS and MVZT, Slovenia;
DST/NRF, South Africa; MICINN, Spain; SRC and Wallenberg Foundation,
Sweden; SER, SNSF and Cantons of Bern and Geneva, Switzerland; NSC, Taiwan;
TAEK, Turkey; STFC, the Royal Society and Leverhulme Trust, United Kingdom;
DOE and NSF, United States of America.

The crucial computing support from all WLCG partners is acknowledged
gratefully, in particular from CERN and the ATLAS Tier-1 facilities at
TRIUMF (Canada), NDGF (Denmark, Norway, Sweden), CC-IN2P3 (France),
KIT/GridKA (Germany), INFN-CNAF (Italy), NL-T1 (Netherlands), PIC (Spain),
ASGC (Taiwan), RAL (UK) and BNL (USA) and from the Tier-2 facilities
worldwide.

\bibliography{pTW_PRD}

\clearpage
\onecolumngrid
\begin{flushleft}
{\Large The ATLAS Collaboration}

\bigskip

G.~Aad$^{\rm 48}$,
B.~Abbott$^{\rm 111}$,
J.~Abdallah$^{\rm 11}$,
A.A.~Abdelalim$^{\rm 49}$,
A.~Abdesselam$^{\rm 118}$,
O.~Abdinov$^{\rm 10}$,
B.~Abi$^{\rm 112}$,
M.~Abolins$^{\rm 88}$,
H.~Abramowicz$^{\rm 153}$,
H.~Abreu$^{\rm 115}$
% ...
\bigskip

% ...
$^{10}$ Institute of Physics, Azerbaijan Academy of Sciences, Baku, Azerbaijan\\
$^{11}$ Institut de F\'isica d'Altes Energies and Departament de F\'isica de la Universitat Aut\`onoma  de Barcelona and ICREA, Barcelona, Spain\\
% ...
$^{48}$ Fakult\"{a}t f\"{u}r Mathematik und Physik, Albert-Ludwigs-Universit\"{a}t, Freiburg i.Br., Germany\\
$^{49}$ Section de Physique, Universit\'e de Gen\`eve, Geneva, Switzerland\\
% ...
$^{88}$ Department of Physics and Astronomy, Michigan State University, East Lansing MI, United States of America\\
% ...
$^{111}$ Homer L. Dodge Department of Physics and Astronomy, University of Oklahoma, Norman OK, United States of America\\
$^{112}$ Department of Physics, Oklahoma State University, Stillwater OK, United States of America\\
% ...
$^{115}$ LAL, Univ. Paris-Sud and CNRS/IN2P3, Orsay, France\\
% ...
$^{118}$ Department of Physics, Oxford University, Oxford, United Kingdom\\
% ...
$^{153}$ Raymond and Beverly Sackler School of Physics and Astronomy, Tel Aviv University, Tel Aviv, Israel\\
% ...
\end{flushleft}


\end{document}